\documentclass[usenatbib]{mnras}
\usepackage{amsmath}
\usepackage{natbib}
\usepackage{epsfig}
\usepackage{graphicx}
\usepackage{amssymb}
\usepackage{url}
\usepackage{xcolor}
\usepackage{soul}

\def\gtsima{$\; \buildrel > \over \sim \;$}
\def\ltsima{$\; \buildrel < \over \sim \;$}
\def\gtrsim{\lower.5ex\hbox{\gtsima}}
\def\lesssim{\lower.5ex\hbox{\ltsima}}

\newcommand{\msun}{M$_{\odot}$}

\definecolor{seagreen}{rgb}{0.190, 0.525, 0.361}

\usepackage{amsmath}	
\usepackage{amssymb}	
\usepackage{mathrsfs}

\begin{document}

\title[IMBHs in YSCs]{Intermediate mass black holes from stellar mergers in young star clusters}

\author[Di Carlo et al.]{\parbox{\linewidth}{Ugo N. Di Carlo$^{1,2,3,4}$\thanks{E-mail: ugoniccolo.dicarlo@unipd.it}, Michela Mapelli$^{1,2,3}$\thanks{E-mail: michela.mapelli@unipd.it}, Mario Pasquato$^{5}$, Sara Rastello$^{1,2}$,}\newauthor{Alessandro Ballone$^{1,2,3}$, Marco Dall'Amico$^{1,2}$, Nicola Giacobbo$^{1,6}$, Giuliano Iorio$^{1,2,3}$,}
\newauthor{Mario Spera$^{1,2,7}$, Stefano Torniamenti$^{1,2,3}$, Francesco Haardt$^{4}$}
\vspace{0.5cm}
\\
$^{1}$Dipartimento di Fisica e Astronomia `G. Galilei', University of Padova, Vicolo dell'Osservatorio 3, I--35122, Padova, Italy
\\
$^{2}$INFN, Sezione di Padova, Via Marzolo 8, I--35131, Padova, Italy
\\
$^{3}$INAF-Osservatorio Astronomico di Padova, Vicolo dell'Osservatorio 5, I--35122, Padova, Italy
\\
$^{4}$Dipartimento di Scienza e Alta Tecnologia, University of Insubria, Via Valleggio 11, I--22100, Como, Italy
\\
$^{5}$ Center for Astro, Particle and Planetary Physics (CAP$^3$), New York University Abu Dhabi
\\
$^{6}$School of Physics and Astronomy, Institute for Gravitational Wave Astronomy, University of Birmingham, Birmingham, B15 2TT, UK
\\
$^{7}$Scuola Internazionale Superiore di Studi Avanzati (SISSA), Via Bonomea 265, I-34136 Trieste, Italy
}
\maketitle \vspace {7cm}
\bibliographystyle{mnras}

\begin{abstract}
Intermediate mass black holes (IMBHs) in the mass range $10^2-10^5\,\mathrm{M_{\odot}}$  
bridge the gap between stellar black holes (BHs) and supermassive BHs. 
 Here, we investigate the possibility that IMBHs form in young star clusters via runaway collisions and BH mergers. We analyze $10^4$ simulations of dense young star clusters, featuring up-to-date stellar wind models and prescriptions for core collapse and (pulsational) pair instability. In our simulations, only 9 IMBHs out of 218 form  via binary BH mergers, with a mass $\sim{}100-140$ M$_\odot$. This channel is strongly suppressed by the low escape velocity of our star clusters. In contrast, IMBHs with masses up to $\sim{}438$ M$_{\odot}$ efficiently form via runaway stellar collisions, especially at low metallicity.  Up to $\sim{}0.2$~\% of all the simulated BHs are IMBHs, depending on progenitor's metallicity. The runaway formation channel is strongly suppressed in metal-rich ($Z=0.02$) star clusters, because of stellar winds. IMBHs are extremely efficient in pairing with other BHs: $\sim{}70$\% of them are members of a binary BH at the end of the simulations. However, we do not find any IMBH--BH merger. More massive star clusters are more efficient in forming IMBHs: $\sim{}8$\% ($\sim{}1$\%) of the simulated clusters with initial mass $10^4-3\times{}10^4$ M$_\odot$  ($10^3-5\times{}10^3$ M$_\odot$) host at least one IMBH.
\end{abstract}


\maketitle

%

\section{Introduction}

Intermediate-mass black holes (IMBHs), with mass in the range $10^2 - 10^5$ M$_\odot$, bridge the gap between stellar-sized black holes (BHs) 
and super-massive BHs. 
IMBHs 
might be the building blocks for super massive BHs in the early Universe \citep[][]{madau2001}, as suggested by the discovery of quasars at redshift as high as $\sim 6$ \citep[e.g.][]{2001AJ....122.2833F}. Currently, we lack unambiguous evidence of IMBH existence from electromagnetic observations, but a number of potential candidates were proposed in the last decades. 
Several globular clusters, including 47~Tucanae \citep{kiziltan2017} and G1 \citep{gebhardt2005}, have been claimed to host IMBHs, based on indirect dynamical evidences \citep{gerssen2002, gebhardt2002, anderson2010,vandermarel2010, lutzgendorf2011,lutzgendorf2013, miller2012, nyland2012,strader2012, lanzoni2013, baumgardt2017,perera2017,lin2018, tremou2018,baumgardt2019,zocchi2019,mann2019}. The hyperluminous X-ray source HLX-1 in the galaxy ESO~243-49 \citep[][]{Farrell2009, godet2014} is possibly the strongest IMBH candidate among ultra-luminous X-ray sources \citep[e.g.,][]{kaaret2001,matsumoto2001,strohmayer2003,vandermarel2004,miller2004,swartz2004,mapelli2010,feng2011,sutton2012,mezcua2015,kaaret2017,lin2020}. Finally, several BHs with mass $\sim{}10^5$ M$_\odot$ might lurk in the nuclei of dwarf galaxies \citep{filippenko1989,filippenko2003,barth2004,greene2004,greene2007,seth2010,dong2012,Kormendy2013, baldassarre2015,Reines2015,mezcua2016,mezcua2018,reines2020}. We point to \cite{mezcua2017} and \cite{greene2020} for two recent reviews.

The gravitational wave event GW190521 \citep{AbbottGW190521, AbbottGW190521astro}, associated to the merger of a binary BH (BBH) with a total mass of $\sim{}150$ M$_\odot$, recently provided strong evidence 
for IMBHs and paved the ground for further observations with current  \citep{LIGOdetector,VIRGOdetector} and future gravitational wave detectors \citep{lisa,ET,cosmicexplorer,kalogera2019,maggiore2020}. 
Moreover, GW190521 confirms the merger formation scenario for IMBHs, where smaller BHs, likely of stellar origin, repeatedly merge in a dense environment, such as a star cluster    \citep[SC, ][]{miller2002,mckernan2012,giersz2015,fishbach2017,gerosa2017,gerosa2019,rodriguez2019,antonini2019,fragione2020,arcasedda2020,mapelli2020,mapelli2021,tagawa2021,tagawa2021b}. Other possible formation scenarios are the collapse of very massive ($>230$ M$_\odot$) metal-poor stars \citep{Bond1984, madau2001,heger2002,schneider2002,spera2017,tanikawa2021}, primordial IMBHs from gravitational instabilities in the  early Universe \citep{2008MNRAS.388.1426K, carr2016,raccanelli2016,sasaki2016,scelfo2018,clesse2020,deluca2021} and runaway collisions of stars in SCs \citep{colgate1967,sanders1970,portegieszwart2002,portegieszwart2004,gurkan2004,freitag2006a,giersz2015,Mapelli2016,dicarlo2019,kremer2020,chon2020,rizzuto2021,das2021}.

Different formation channels may be at work in different environments \citep{volonteri2010}, with SCs representing the ideal place both for repeated mergers of stellar-mass BHs \citep[e.g.,][]{miller2002} and for the formation of very massive stars through runaway collisions in the early stages of the cluster's evolution \citep[e.g.,][]{fujii2013}. 
The internal dynamics of SCs is expected to play a key role in modulating the efficiency of the BBH merger scenario, with core collapse being closely tied to increased dynamical interactions \citep[][]{spitzer1987,goodman1989} and to the dynamical hardening of binary systems  \citep[][]{1983MNRAS.204P..19S, goodman1993,trenti2007,2012MNRAS.425.2872H, 2019ApJ...876...87B, 2021arXiv210302424D}.
The main advantage of the repeated merger mechanism is that it works for the entire lifetime of a SC: the IMBH can assemble and grow as long as the SC survives \citep{giersz2015}. 
However, a merger chain can come to an abrupt end if the BBH merger product is ejected from the SC due to gravitational-wave recoil \citep{merritt2004,madau2004,holley2008}. The magnitude of such kick can largely exceed the escape velocity of the host SC \citep{favata2004,lousto2011}, thus the hierarchical build up of an IMBH via repeated BBH mergers is more likely to take place in the most massive SCs, like nuclear SCs, because of their large escape velocity \citep{antonini2019,fragione2020,mapelli2021}.

The runaway collision mechanism, on the other hand, can be effective even in less massive SCs, like young SCs. The most massive stars in a young SC are more likely to undergo collisions than lighter stars, because dynamical friction quickly brings them to the core \citep[][]{gaburov2008b}.
The high central density of young SCs, which is further enhanced in the first Myr due to gravothermal collapse, greatly favours stellar collisions \citep{freitag2006b,portegieszwart2010}, which help to build up very massive stars that may collapse into IMBHs.
The main issue is that very massive stars lose a conspicuous portion of their mass through stellar winds \citep[e.g.][]{vink2011}. The IMBH can form only if the star preserves enough mass, so this mechanism is more efficient at lower metallicity, where stellar winds are less powerful \citep{Mapelli2016}.
Moreover, all the collisions need to take place before the massive stars in the SC turn into BHs, which happens in few Myr.

Despite being less massive than globular clusters and nuclear SCs, young SCs make up the vast majority of SCs in the Universe \citep{kroupa2002}, and their cumulative contribution to IMBH statistics may thus be significant. 
In this paper, we study the demography of IMBHs in young SCs through $N$-body simulations, including up-to-date stellar wind models and a treatment for (pulsational) pair instability in massive and very massive stars. We present a large set of $10^4$ state of the art direct $N$-body simulations with fractal initial conditions \citep[to mimic the clumpiness of star forming regions, e.g.][]{goodwin2004} and with a large initial binary fraction ($f_{\rm{bin}}=1$ for stars with mass $>5$ M$_\odot$). 

\section{Methods}

\begin{table}
\begin{center}
\caption{\label{tab:table1} Initial conditions.} \leavevmode
\begin{tabular}[!h]{cccccc}
\hline
Set     & $Z$ & $N_{\rm sim}$ & $r_{\rm h}$  & $D$ & ref. \\
\hline
D2019HF &0.002 & 2000   & M12 & 2.3 & D19 \\
\hline
D2019LF &0.002 & 2000   & M12 & 1.6 & D19 \\
\hline
D2020A &0.02  & 1000   & M12   & 1.6 & D20\\
        &0.002 & 1000   & M12   & 1.6 & D20\\
        &0.0002 & 1000  & M12   & 1.6 & D20 \\
\hline
D2020B &0.02  & 1000   & 1.5 pc  & 1.6 & D20\\
        &0.002 & 1000   & 1.5 pc  & 1.6 & D20\\
        &0.0002 & 1000  & 1.5 pc  & 1.6 & D20 \\
\hline
\end{tabular}
\end{center}
\begin{flushleft}
\footnotesize{Column~1: name of the simulation set; column~2: metallicity $Z$; column~3: number of runs performed per each set; column~4: half-mass radius $r_{\rm h}$. M12 indicates that half-mass radii have been drawn according to \protect{\cite{markskroupa12}}. Column~5: fractal dimension ($D$);  column~6: reference for each simulation set. D19 and D20 correspond to \protect{\cite{dicarlo2019}} and \protect{\cite{dicarlo2020}}, respectively.}
\end{flushleft}
\end{table}

The simulations discussed in this paper were performed using the same codes and methodology described in \cite{dicarlo2019}. We use the direct summation $N$-Body code \textsc{nbody6++gpu} \citep{wang2015}, which we coupled with the population synthesis code \textsc{mobse} \citep{mapelli2017,giacobbo2018}. {\sc mobse} includes up-to-date prescriptions for massive stellar winds \citep{giacobbo2018b}, electron-capture supernovae \citep{giacobbo2018c}, core-collapse supernovae \citep{fryer2012}, natal kicks \citep{giacobbo2020} and (pulsational) pair instability supernovae \citep{mapelli2020b}. Orbital decay induced by gravitational wave emission is treated as described in \cite{peters1964}. 
If a star merges with a BH or with a neutron star, \textsc{mobse} assumes that the entire mass of the star is lost by the system and not accreted by the compact object.

\begin{figure}
  \center{
    \epsfig{figure=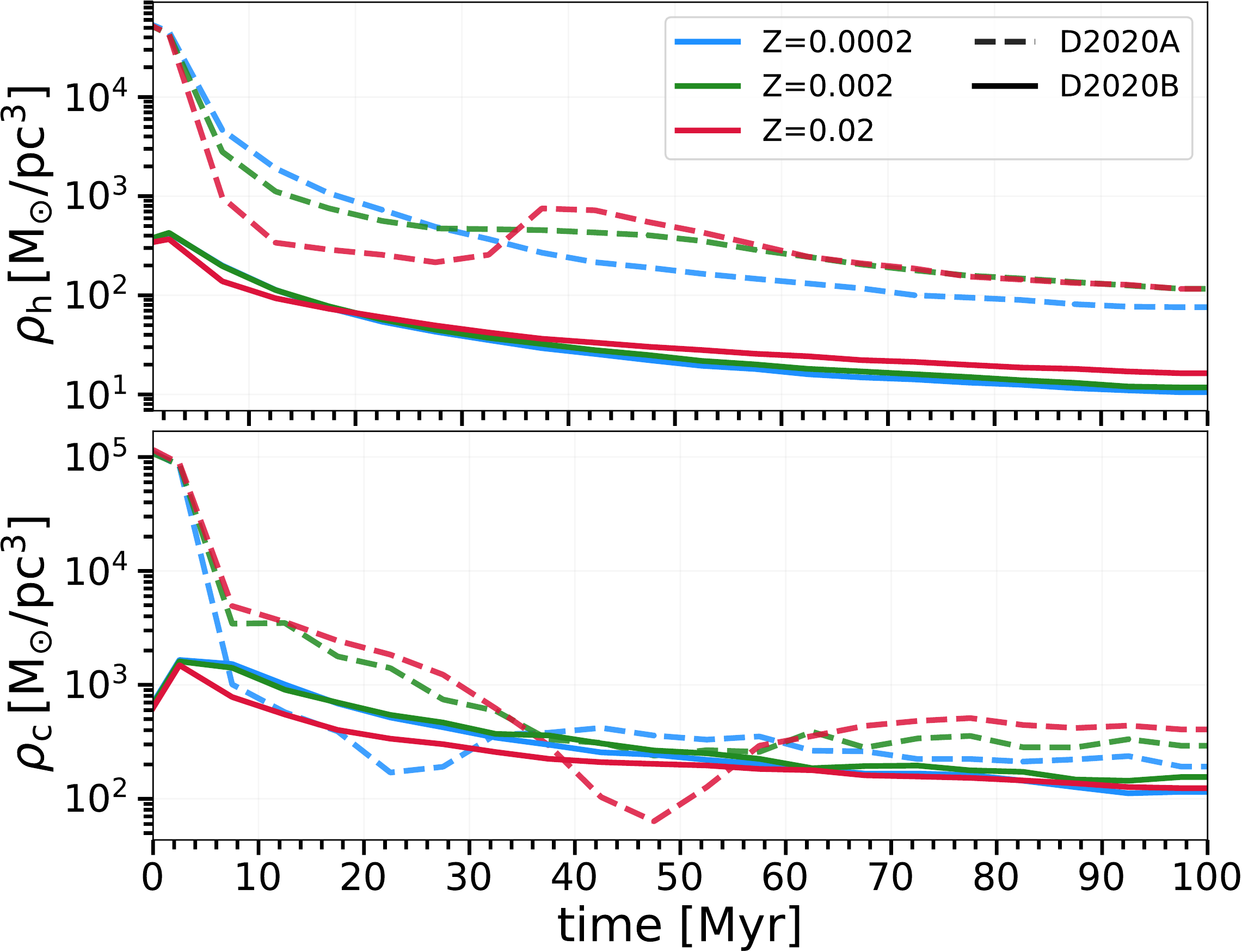,width=8.0cm}
    \caption{\label{fig:dens} Evolution of the median of the mass density at the half mass radius ($\rho_{\rm{h}}$, top panel) and at the core radius ($\rho_{\rm{c}}$, bottom panel) for the SCs in sets D2020A ($r_{\rm h}$ from \citealt{markskroupa12}, dashed lines) and D2020B ($r_{\rm h}=1.5$ pc, solid lines) at different metallicities.}}
\end{figure}

In this work, we have analyzed $10^4$ simulations of young SCs; 4000 of them are the simulations presented in \cite{dicarlo2019}, while the remaining 6000 are the ones discussed in \cite{dicarlo2020}. The initial conditions of the simulations are summarized in Table~\ref{tab:table1}. Young SCs are asymmetric, clumpy systems. Thus, we model them with fractal initial conditions \citep{kuepper2011}, to mimic the clumpiness of stellar forming regions \citep{goodwin2004}. The level of fractality is decided by the parameter $D$, where $D=3$ means homogeneous distribution of stars. In this work, we assume $D=1.6$ in most of our simulations (8000 runs). We also probe a lower degree of fractality ($D=2.3$) in the remaining 2000 runs. 
The total initial mass $M_{\rm SC}$ of each SC ranges from $10^3$ \msun{} to $3\times{}10^4$ \msun{} and it is drawn from a distribution $dN/dM_{\rm SC}\propto M_{\rm SC}^{-2}$, as the SC initial mass function described in \cite{lada2003}. Thus, the mass distribution of our simulated SCs mimics the mass distribution of SCs in Milky Way-like galaxies. We choose the initial half-mass radius $r_{\rm h}$ according to  \cite{markskroupa12} in 7000 simulations, and we adopt a fixed value $r_{\rm h}=1.5$ pc for the remaining 3000 simulations. These different choices of $r_{\rm h}$ have a strong impact on the density of the SCs, as shown in Figure \ref{fig:dens}.

We generated the initial conditions with {\sc mcluster} \citep{kuepper2011}. The initial masses of the stars follow a \cite{kroupa2001} initial mass function, with minimum mass 0.1 \msun{}  and maximum mass 150 \msun{}. We set an initial total binary fraction of $f_{\mathrm{bin}}=0.4$. Since {\sc mcluster} pairs up the most massive primary stars first, according to a distribution
$\mathcal{P}(q)\propto{}q^{-0.1}$ (where $q=m_2/m_1\in{}[0.1,1]$, \citealt{sana2012}), all the stars with mass
$\,m\,\ge{}5$~M$_\odot$ are  members of binary systems, while stars with mass
$m\,<\,5$~M$_\odot$ are randomly paired until a total binary fraction
$f_{\mathrm{bin}}=0.4$ is reached. This procedure results in a mass-dependent initial binary fraction consistent with the multiplicity properties of O/B-type stars (\citealt{sana2012,moe2017}). The orbital periods and eccentricities are also drawn from the distributions described in \cite{sana2012}.  

 We simulate SCs with three different metallicities: $Z=0.0002,$ 0.002 and 0.02 (approximately 1/100, 1/10 and 1 Z$_\odot$, assuming $Z_{\odot}=0.02$ from \citealt{anders1989}). Each SC is simulated for 100 Myr in a solar neighbourhood-like static external tidal field \citep{wang2016}. At 100 Myr, we extract the BBHs which are still present in our simulations and we integrate their semi-major axis and eccentricity evolution by gravitational wave emission up to a Hubble time (13.6 Gyr), according to \cite{peters1964}.

 The main differences between the simulations in \cite{dicarlo2019} and \cite{dicarlo2020} are the efficiency of common envelope ejection ($\alpha=3$ in \citealt{dicarlo2019} and $\alpha{}=5$ in \citealt{dicarlo2020}), and the chosen model of core-collapse supernova (the rapid and the delayed models from \citealt{fryer2012} are adopted in \citealt{dicarlo2019} and in \citealt{dicarlo2020}, respectively). 
The population of IMBHs is not strongly affected by these differences.
We divide the simulations in four different sets whose names and properties are shown in Table~\ref{tab:table1}. We refer to \cite{dicarlo2019} and \cite{dicarlo2020} for further details on the code and on the simulations.

\section{Results}

\begin{table}
\begin{center}
\caption{\label{tab:scimbh} Percentage of SCs that form 0, 1 and 2 IMBHs per different metallicity, initial SC mass and simulation set.} \leavevmode
\begin{tabular}[!h]{cccc}
\hline
Set & 0 IMBHs & 1 IMBH & 2 IMBHs\\
\hline
All & 97.97 \% & 1.97 \% &  0.06 \% \\
\hline
$Z=0.02$ & 99.85 \% & 0.15 \% &  0 \% \\
$Z=0.002$ & 97.81 \% & 2.17 \% & 0.02 \% \\
$Z=0.0002$ & 96.5 \% & 3.25 \% & 0.25 \% \\
\hline
$10^3\leq M_{\rm{SC}}/$M$_\odot<5\times{}10^3$     & 99.01 \% & 0.99 \% &  0 \% \\
$5\times{}10^3\leq M_{\rm{SC}}/$M$_\odot<10^4$     & 95.73 \% & 4.12 \% &  0.15 \% \\
$10^4\leq M_{\rm{SC}}/$M$_\odot\leq 3\times{}10^4$ & 92.13 \% & 7.42 \% &  0.45 \% \\
\hline
D2019HF & 96.90 \% & 3.05 \% & 0.05 \% \\
D2019LF & 98.2 \% & 1.80 \% & 0.0 \% \\
\hline
D2020A & 98.03 \% & 1.94 \% & 0.03 \% \\
D2020B & 98.47 \% & 1.40 \% & 0.13 \% \\
\hline

\end{tabular}
\end{center}
\begin{flushleft}
\footnotesize{Column~1: Simulation set; column~2: percentage of SCs that form no IMBHs; column~3: percentage of SCs that form 1 IMBH; column~4: percentage of SCs that form 2 IMBHs.}
\end{flushleft}
\end{table}


\begin{figure}
  \center{
    \epsfig{figure=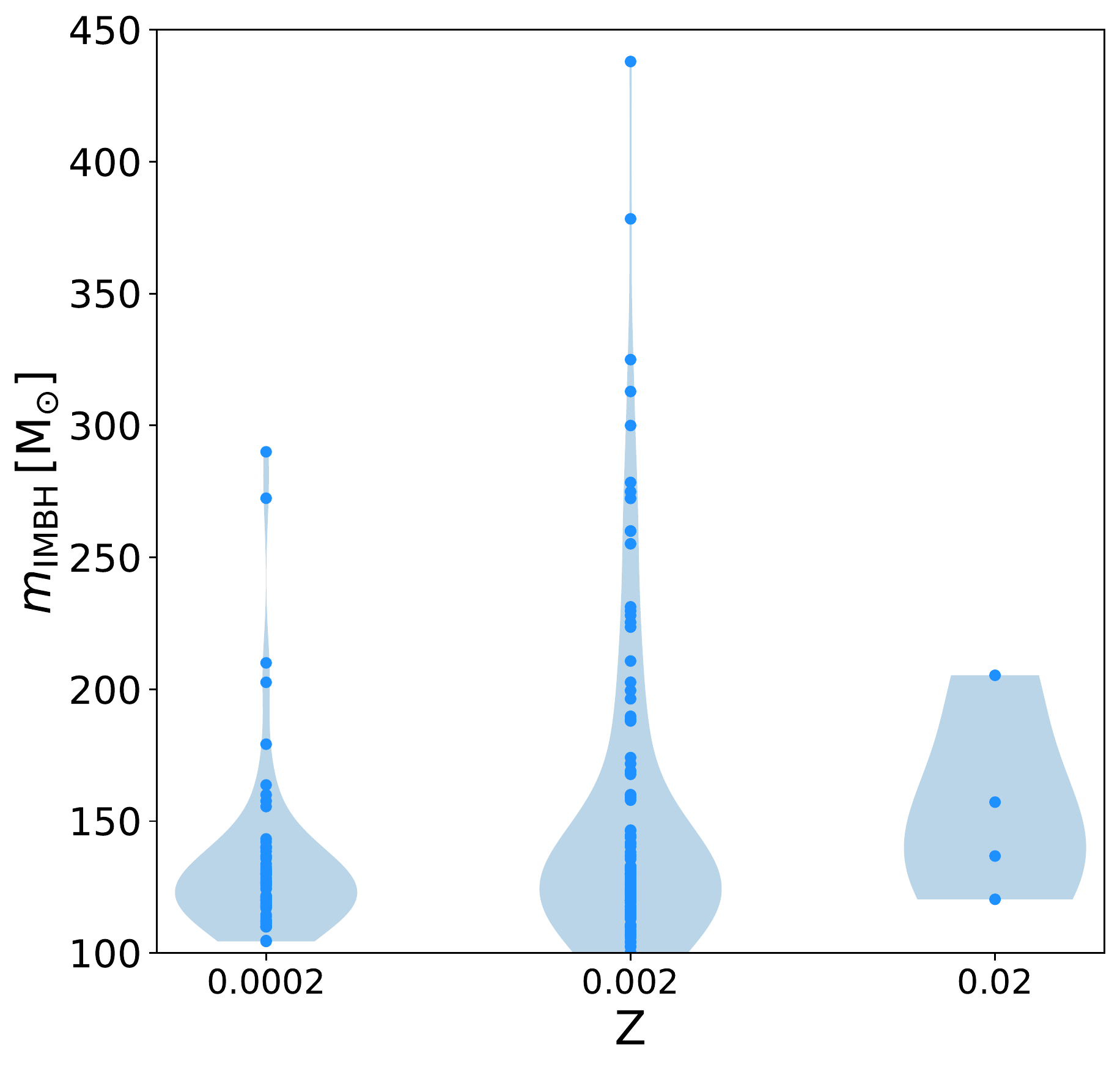,width=8.0cm}
    \caption{\label{fig:m_z_violin} Distributions of IMBH masses for each metallicity. Blue filled circles mark the values of the masses, while the horizontal extent of each light blue region (violin plot) is proportional to the number of IMBHs at a given mass value.}}
\end{figure}


\begin{figure}
  \center{
    \includegraphics[width = 0.45\textwidth]{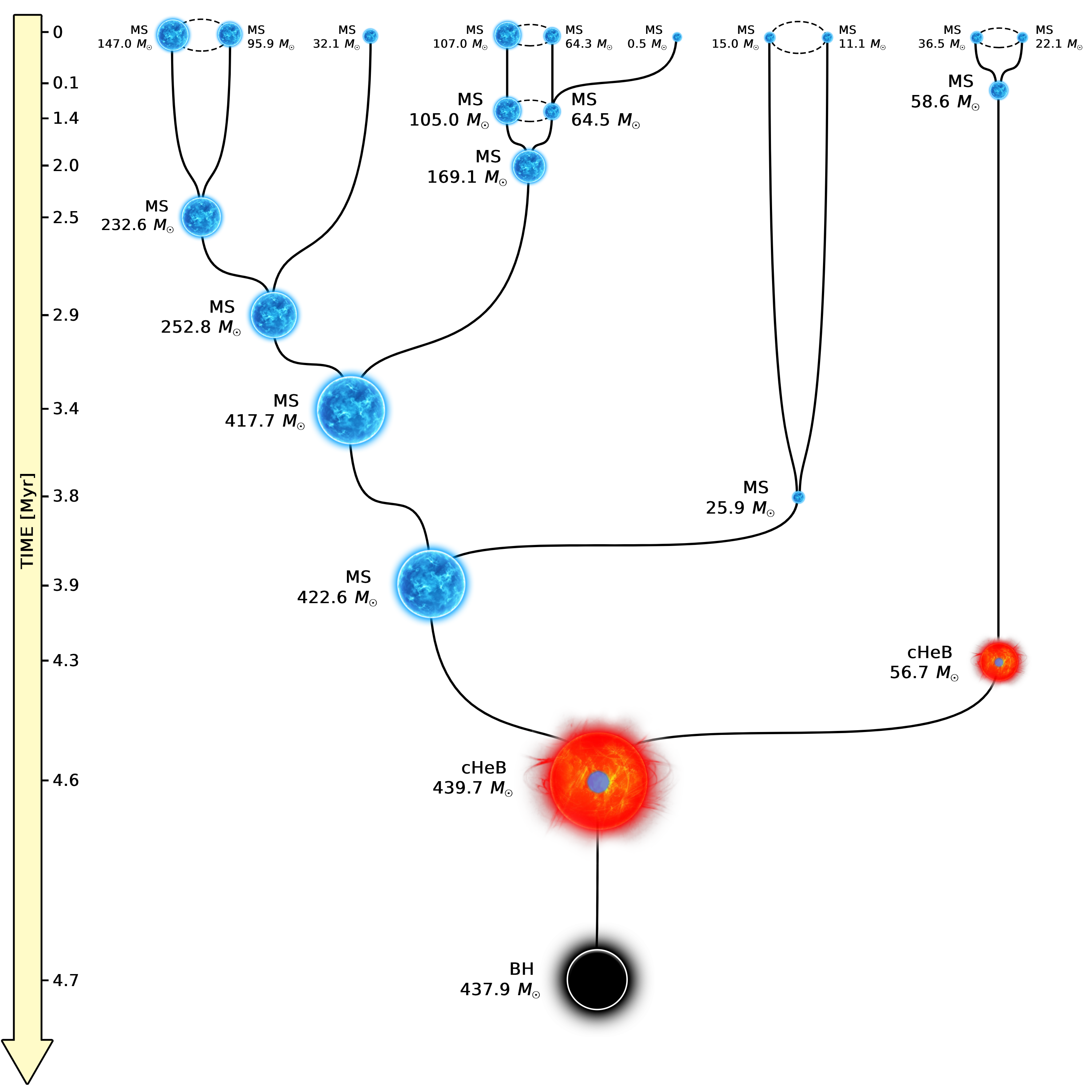}
    \caption{{\label{fig:mostmassive} Formation history of the most massive IMBH found in our simulations. Main sequence stars (with label MS) are represented as blue stars;  core helium burning stars (label cHeB) are visualized as red stars with a blue core; black holes (label BH) are shown as black circles. The mass of each object is shown next to them. The time axis and the size of the objects are not to scale. The formation of this IMBH takes place in a SC with metallicity $Z=0.002$.}}}
\end{figure}

\begin{figure}
  \center{
    \epsfig{figure=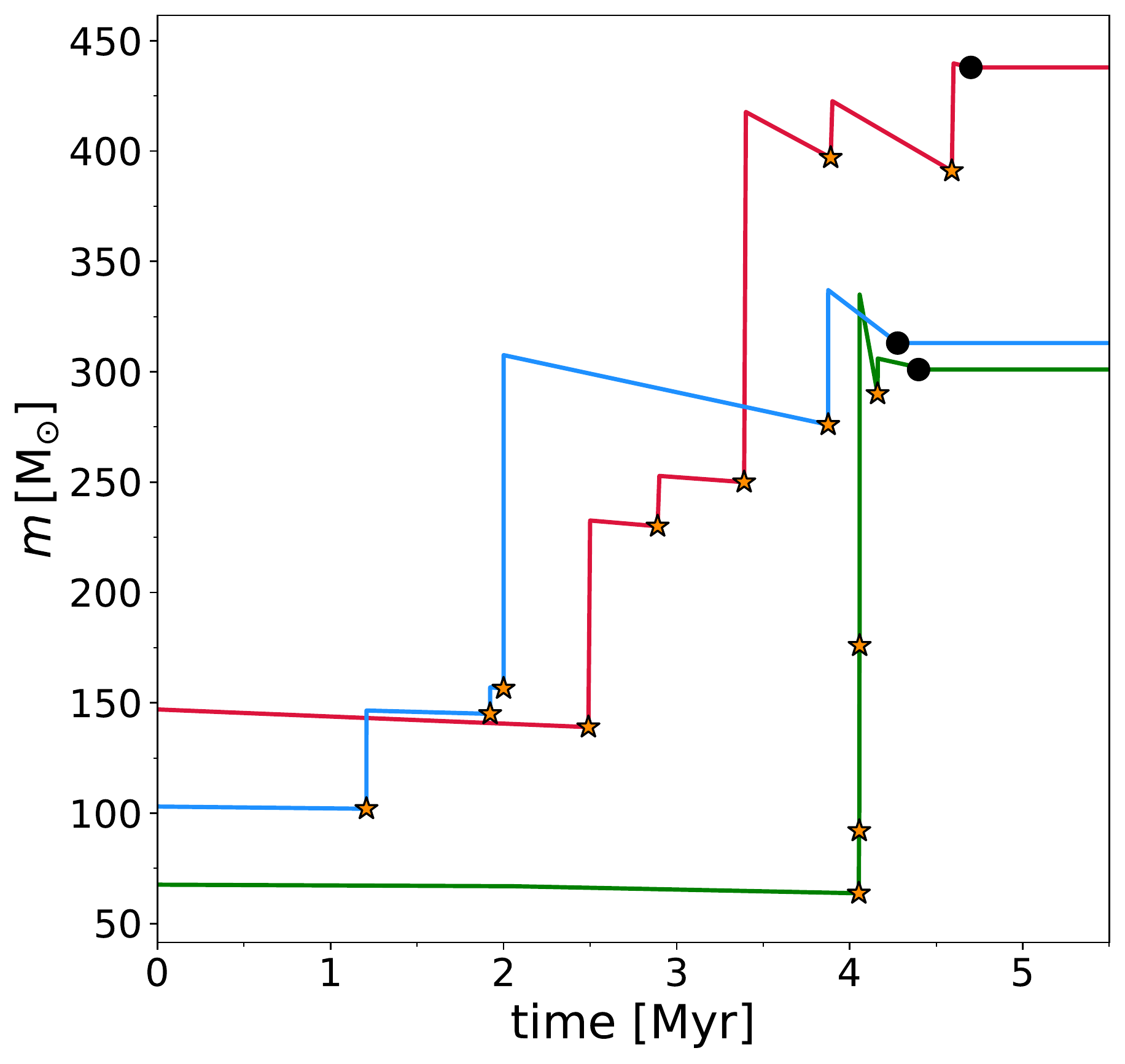,width=8.0cm}
    \caption{{\label{fig:massev} Mass evolution of three stars which end up forming an IMBH. Yellow stars mark stellar mergers. Black filled circles indicate the moment when the IMBH form.}}}
\end{figure}

\subsection{Formation of IMBHs}

We find a total of 218 IMBHs, which make up only $\sim0.2\%$ of all the BHs formed in our simulations.
The formation of IMBHs is a rare event in our simulations: as shown in Table~\ref{tab:scimbh}, only $1.97\%$ of all the simulated SCs form one IMBH, and only $0.06\%$ of them form two IMBHs. 
The vast majority of the simulated  IMBHs (209) form via the runaway collision mechanism, while only 9 IMBHs form via BBH merger. This is expected, because the low escape velocity ($\lesssim10$~km~s$^{-1}$) of our simulated SCs makes the BBH merger scenario less efficient.

Figure~\ref{fig:m_z_violin} shows the masses of all the formed IMBHs for each metallicity. IMBHs form with mass up to $\sim438$~\msun, but $\sim78\%$ of all the formed IMBHs has a mass between $100$~\msun{} and $150$~\msun. Less massive IMBHs are more likely to form.
A schematic formation history of the most massive IMBH is shown in Figure \ref{fig:mostmassive}; this IMBH forms via runaway collision: a total of 10 stars participate to the formation of a very massive star which promptly undergoes direct collapse to form the IMBH. All the other IMBHs which form via runaway collision follow a similar pattern, involving multiple stellar mergers in the first few Myr of our simulations. The mass evolution of three stars which end up forming an IMBH is shown in Figure~\ref{fig:massev}. We see that mass growth via stellar mergers overcomes stellar wind mass loss. The formation of the IMBHs via runaway collision always occurs within the first $\sim 5$ Myr from the beginning of the simulation.

As expected, IMBH formation is much less efficient at higher metallicity because stellar winds are more powerful; only four IMBHs form at solar metallicity. The fraction of IMBHs with respect to the total number of BHs per each metallicity is $\sim{}0.4$~\%, $\sim{}0.2$~\% and $\sim{}0.02$~\% for $Z=0.0002$, 0.002 and 0.02, respectively. Table~\ref{tab:scimbh} shows that the percentage of simulated SCs which form at least one IMBH, grows as the metallicity decreases, going from $0.15\%$ at $Z=0.02$ to $3.5\%$ at $Z=0.0002$. 
We obtained a larger absolute number of IMBHs at metallicity $Z=0.002$ (Fig.~\ref{fig:m_z_violin}), only because we ran three times more simulations with $Z=0.002$ than with either $Z=0.02$ or $Z=0.0002$ (Table~\ref{tab:table1}). We performed several statistical tests to check if the fact that the most massive IMBHs form at $Z=0.002$ is statistically significant or is just another effect of the larger number of simulation at this metallicity. Our tests suggest that intermediate-metallicity SCs (with $Z=0.002=0.1$ Z$_\odot$) tend to form more massive IMBHs than metal-poor SCs (with $Z=0.0002=0.01$ Z$_\odot$), even if there is no conclusive evidence for this result (see Appendix~\ref{sec:statapp} for details). This possible difference could be interpreted as the interplay between relatively inefficient mass loss and large maximum stellar radii \citep{chen2015}: at $Z=0.002$ stellar winds are still rather inefficient (as for lower metallicity), but massive stars can develop very large radii (in the terminal-age main sequence and the giant phase), enhancing the probability of collisions with other stars.



\subsection{Distance from centre of mass}

\begin{figure}
  \center{
    \epsfig{figure=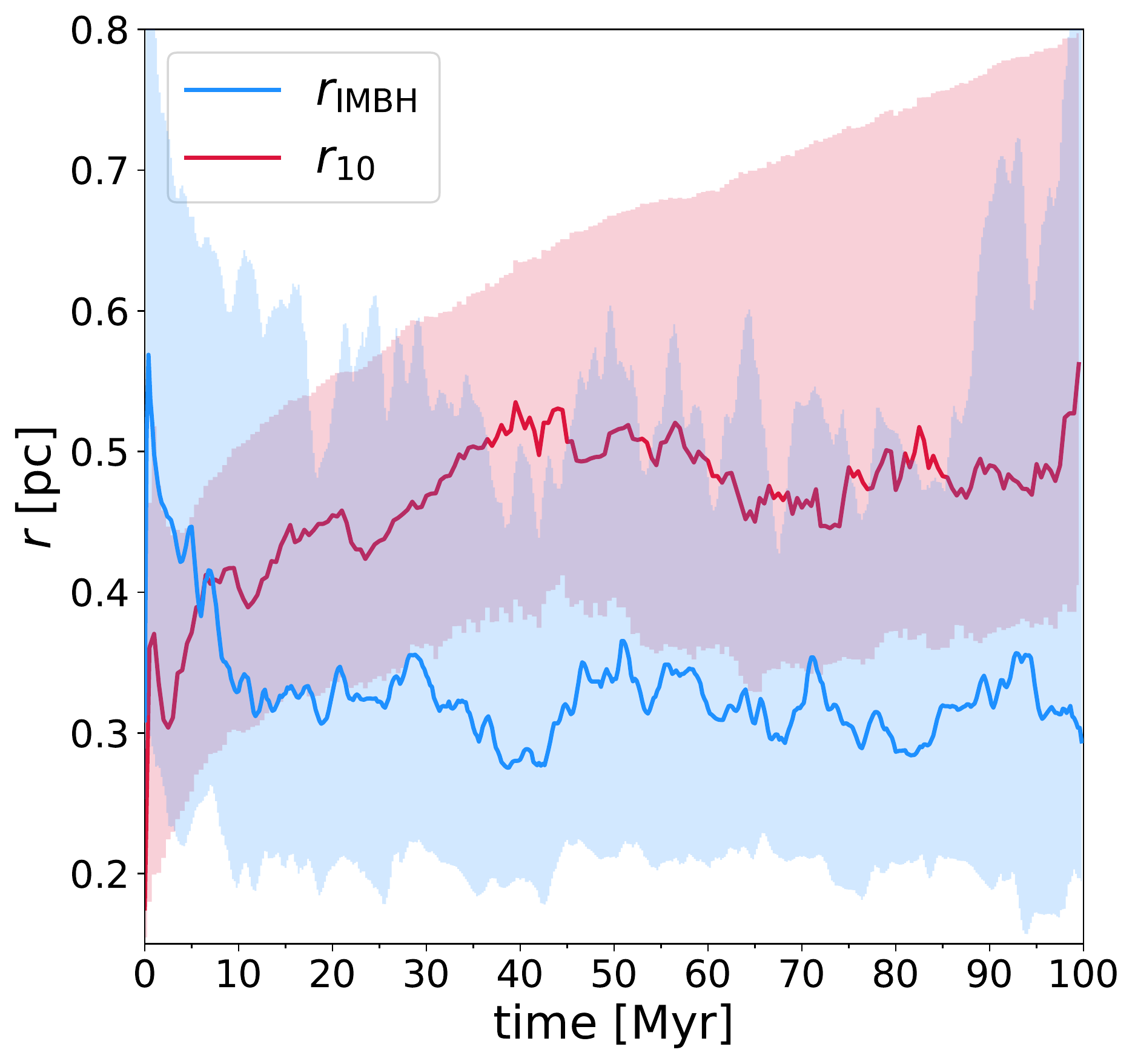,width=8.0cm}
    \caption{\label{fig:distance}Evolution of $r_{\mathrm{IMBH}}$ (distance of the IMBHs from the centre of mass of the SC). The blue line shows the median of the values of $r_{\mathrm{IMBH}}$, while the red line is the median of the $10$\% Lagrangian radii ($r_{10}$) of the host SCs. The filled areas represent the 25th$-$75th percentile confidence intervals. Only IMBHs which do not escape from their host SC are shown in this Figure. A simple moving average over 5 time-steps has been performed to filter out statistical fluctuations.}}
\end{figure}

Figure~\ref{fig:distance} shows the evolution of the median distance of the IMBHs with respect to the centre of mass of the host SCs. In order to undergo a sufficient number of stellar mergers, we expect the progenitor stars of IMBHs to lie in the densest central regions of the SC. In our simulations, we find that this is not completely the case: IMBH progenitors (i.e. before $t\sim 5\,\mathrm{Myr}$) tend to lie outside the $10\%$ Lagrangian radius. This is likely a consequence of the fractal initial conditions, because the IMBH formation may take place in a clump far away from the centre of mass of the SC, before the mergers between the clumps take place.
After the IMBHs form, they tend to rapidly sink towards the centre of the SC and to remain there until the end of the simulation.

\subsection{Escaped IMBHs}

Figure~\ref{fig:distance} shows only IMBHs which do not escape from their host SC. In our simulations, particles escape from the SC when their distance from the center of mass becomes larger than twice the tidal radius. We find that $\sim 54.2\%$ of all the formed IMBHs are ejected or evaporate from the SC in the first 100 Myr.
The percentage of IMBHs with mass $m_{\mathrm{IMBH}}<150$ \msun that escape from their host SC is $\sim57.0\%$, and only $\sim31.8\%$ for IMBHs with $m_{\mathrm{IMBH}}\geq 150$ \msun. This means that the most massive IMBHs are more likely to be retained inside their host SC.

\begin{figure}
  \center{
    \epsfig{figure=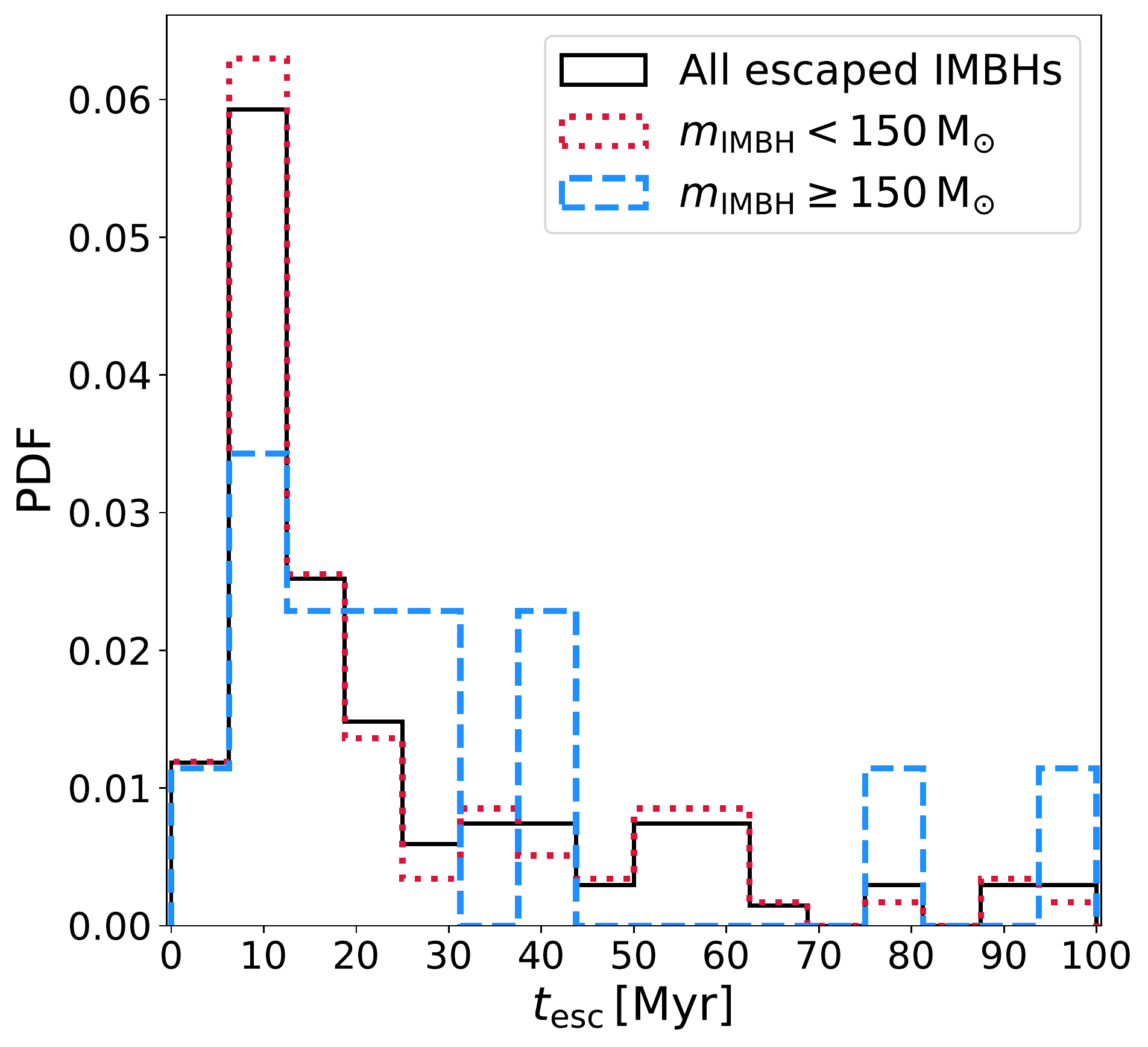,width=8.0cm}
    \caption{\label{fig:esctimes_mimbh} Probability distribution function of the escape time $t_{\mathrm{esc}}$ of IMBHs. All the ejected IMBHs are shown by the solid black line. IMBHs with mass $m_{\mathrm{IMBH}}<150$ \msun{} are shown by the dotted red line, while IMBHs with mass $m_{\mathrm{IMBH}}\geq150$ \msun{} are shown by the dashed blue line.}}
\end{figure}

\begin{figure}
  \center{
    \epsfig{figure=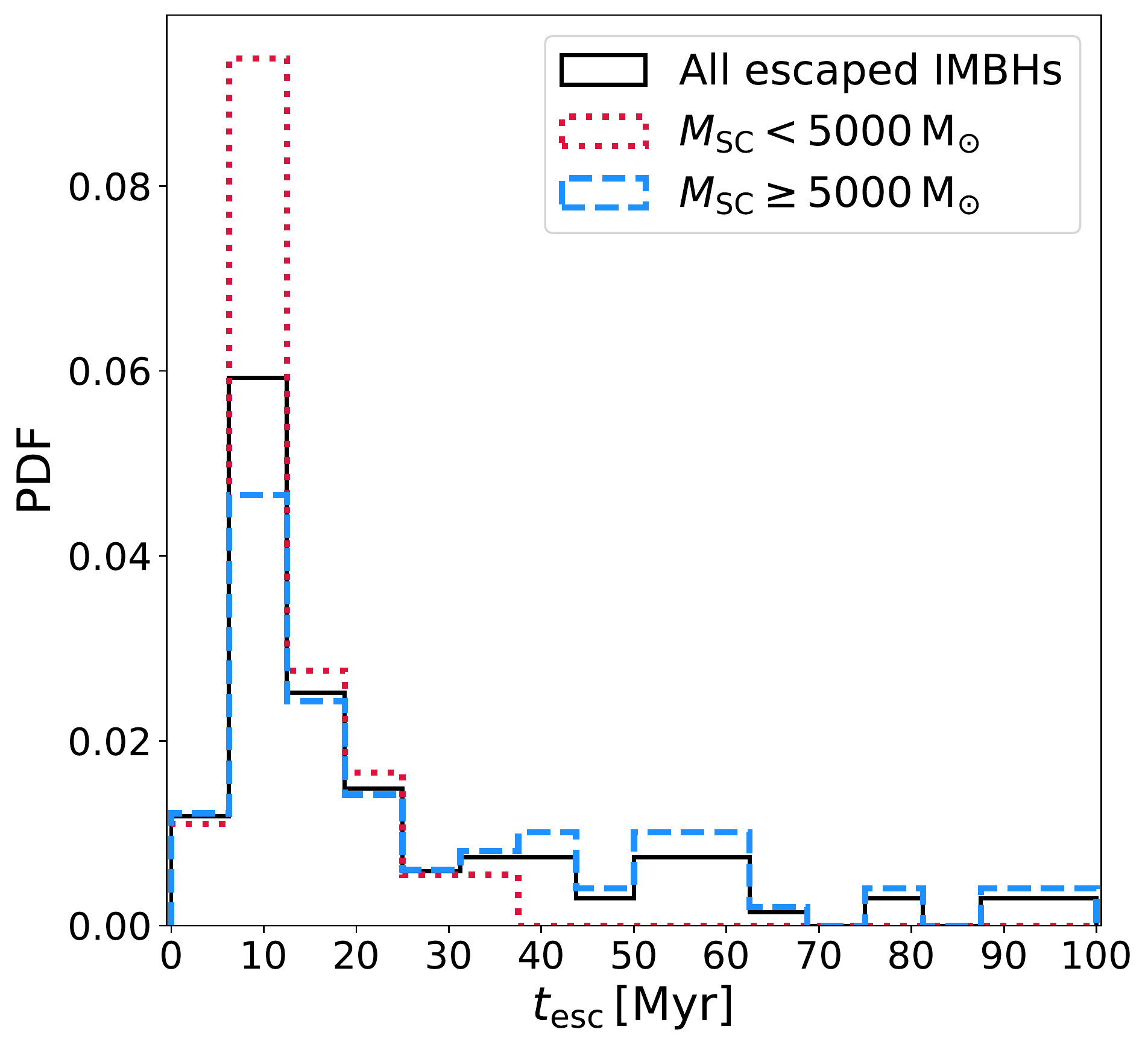,width=8.0cm}
    \caption{\label{fig:esctimes_msc} Probability distribution function of the escape time $t_{\mathrm{esc}}$ of the IMBHs. All the ejected IMBHs are shown by the solid black line. IMBHs formed in SCs with mass $M_{\mathrm{SC}}<5000$ \msun{} are shown by the dotted red line, while IMBHs formed in SCs with mass $M_{\mathrm{SC}}\geq5000$ \msun{} are shown by the dashed blue line.}}
\end{figure}

Figure~\ref{fig:esctimes_mimbh} shows the distribution of the escape times $t_{\mathrm{esc}}$ of ejected IMBHs. All IMBHs and especially the less massive ones tend to be ejected within the first $25$ Myr, with a strong peak between $5$ and $15$ Myr. More massive IMBHs tend to escape at later times with respect to the less massive ones. Figure \ref{fig:esctimes_msc} distinguishes between IMBHs that formed in SCs with initial mass smaller and larger than $5\times{}10^3$ \msun. Both distributions peak around $10$ Myr, but all the ejected IMBHs which form in the less massive SCs escape within the first $40$ Myr, while IMBHs born in more massive SCs may escape at later times. 

We find that the percentages of escaped IMBHs which form in SCs with initial mass $M_{\mathrm{SC}}<5000$ \msun{} and $M_{\mathrm{SC}}\geq 5000$ \msun{}  are $37.7\%$ and $59.8\%$, respectively. This means that IMBHs are more likely to be ejected from more massive SCs. This may seem surprising, because we expect more massive SCs to retain more IMBHs due to their larger escape velocities. Massive SCs, however, have denser cores where dynamical interactions that may eject IMBHs are more frequent.

\subsection{IMBHs in binary systems}

\begin{figure}
  \center{
    \epsfig{figure=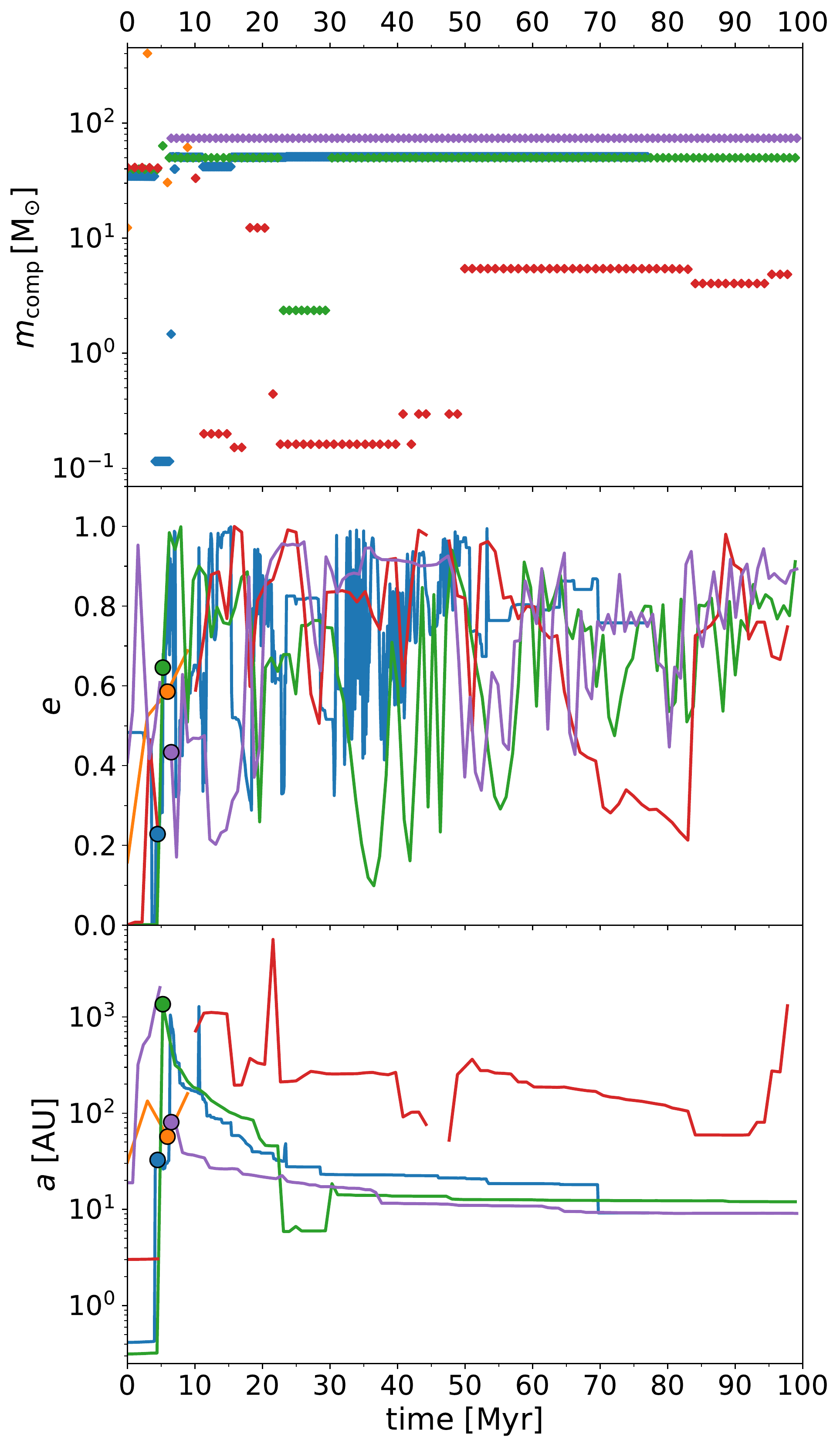,width=8.0cm}
    \caption{\label{fig:a_ecc_mcomp}Evolution of the mass of the  IMBH companion $m_{\mathrm{comp}}$ (upper panel), eccentricity $e$ (middle panel) and semi-major axis $a$ (lower panel) of the binaries which contain the five most massive IMBHs formed in our simulations. Coloured filled circles represent the formation time of the IMBH, i.e. the time when the progenitor star becomes an IMBH. The red circle is missing because the progenitor of this IMBH is single when it collapses to a BH.}}
\end{figure}

IMBHs are the most massive objects which form in our simulations, and therefore they are  likely to interact with other stars and form binary systems \citep{blecha2006,macleod2016}. In our simulations, IMBHs spend  $\sim85\%$ of the time,  on average, being part of a binary or triple system. Figure \ref{fig:a_ecc_mcomp} shows the evolution of the companion mass $m_{\mathrm{comp}}$, orbital eccentricity $e$ and semi-major axis $a$  for the five most massive IMBHs that form in our simulations.  All these five massive IMBHs spend almost all their time in binary systems, except for one of them, which is ejected from the SC after $\sim 10$ Myr.

 These binaries are subject to continuous dynamical interactions which perturb their orbital properties. Eccentricity wildly oscillates and tends to assume values between $0.6$ and $0.9$ (Figure~\ref{fig:a_ecc_mcomp}). Each variation of eccentricity corresponds to a smaller variation of the semi-major axis, which tends to shrink over time. At the end of the simulation, two of the IMBHs shown in Figure~\ref{fig:a_ecc_mcomp} are still members of a binary system with a companion BH. We have re-simulated the host SCs of these IMBHs for up to $1$ Gyr to check for possible mergers, but none of these  binary systems does merge\footnote{{\sc nbody6++gpu} does not include a treatment for post-Newtonian terms. We cannot exclude the possibility that dynamical interactions combined with a treatment of post-Newtonian terms might allow the merger of some of our IMBH--BH binaries.}. 

\begin{figure}
  \center{
    \epsfig{figure=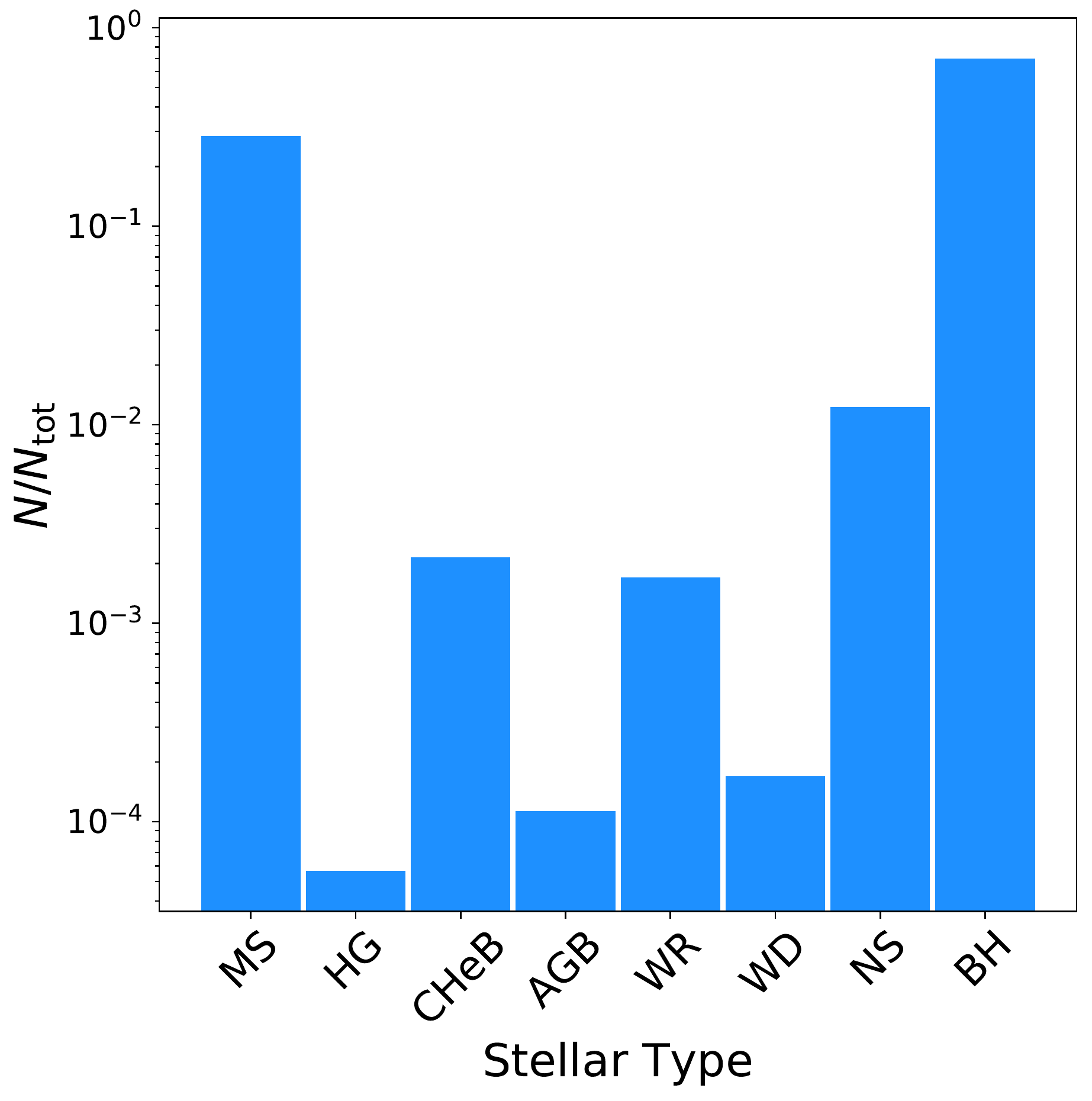,width=8.0cm}
    \caption{\label{fig:kdist}Distribution of the stellar types of the companions of all the simulated IMBHs throughout the entire simulation ($0-100$ Myr). The values on the $y-$axis are normalized to the total number of companions $N_{\rm tot}$. The represented stellar types are: main sequence (MS); Hertzsprung gap (HG); core helium burning (CHeB); asymptotic giant branch (AGB); Wolf-Rayet (WR); white dwarf (WD); neutron star (NS) and black hole (BH).}}
\end{figure}

The simulated IMBHs tend to couple with other massive  members of the SC, because dynamical exchanges generally lead to the build up of more and more massive binaries 
\citep{hills1980}. Figure \ref{fig:kdist} shows the distribution of the stellar types of the IMBH's companions. Besides main sequence stars, which are the most common stellar type in the simulated SCs, we see that the most common binary companions of IMBHs are BHs and neutron stars. 

\subsubsection{IMBH--BH binaries}

\begin{figure}
  \center{
    \epsfig{figure=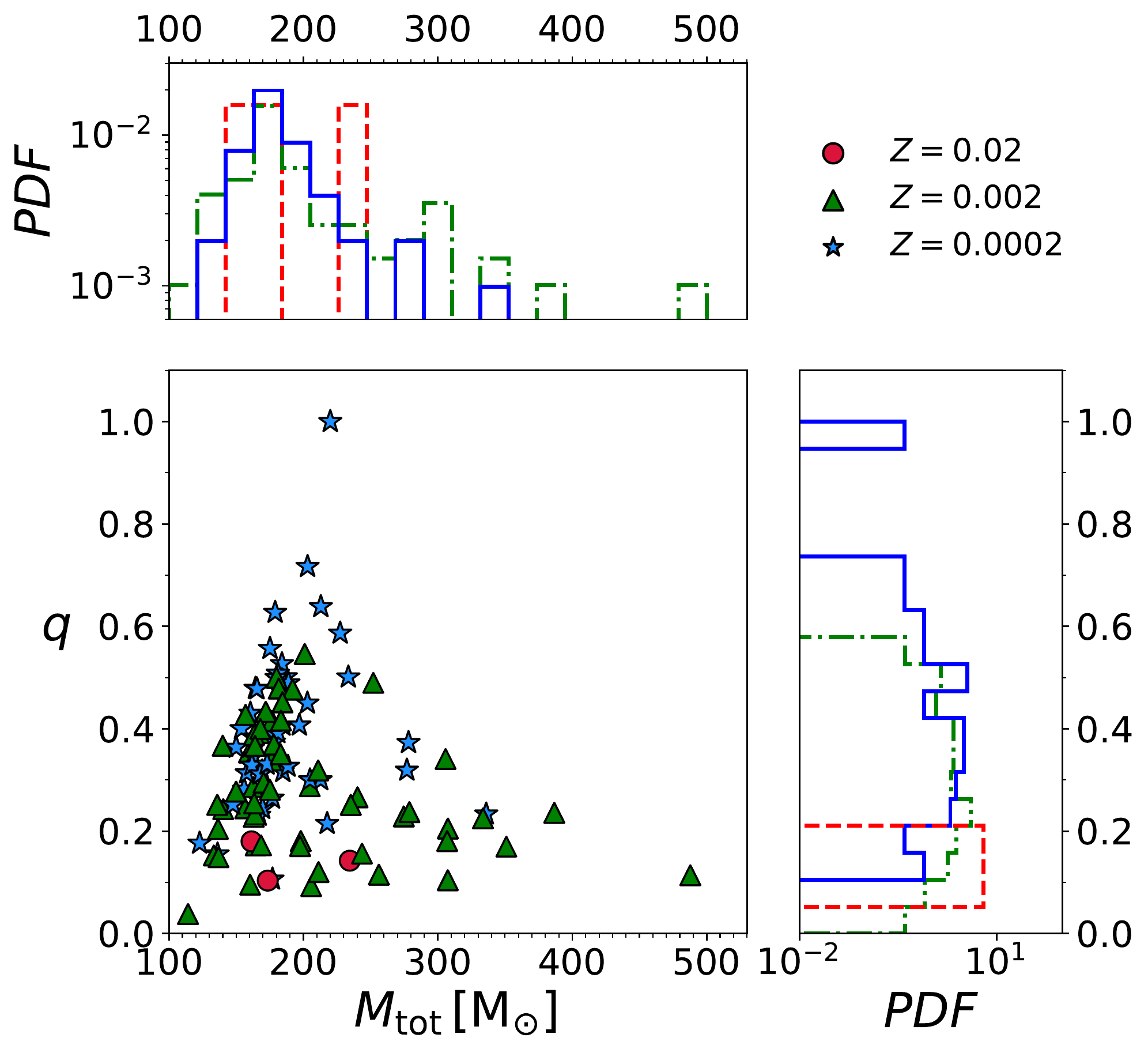,width=8.0cm}
    \caption{\label{fig:qmtotimbh} Mass ratio $q=m_2/m_1$ versus total mass $M_{\rm tot}=m_1+m_2$ of IMBH--BH binaries. Circles, triangles and stars refer to $Z=0.02$, 0.002 and 0.0002, respectively. The marginal histograms show the distribution of $q$ (on the $y-$axis) and $M_{\rm tot}$ (on the $x-$axis).
    Solid blue, dot-dashed green and dashed red histograms refer to $Z=0.0002,$ 0.002 and 0.02, respectively.}}
\end{figure}

We analyze the properties of all BBHs which host at least one IMBH in our simulations (hereafter,  IMBH--BH binaries).
From Figure \ref{fig:kdist}, it is apparent that IMBHs are extremely efficient in finding a companion BH: $\sim70\%$ of all IMBHs reside in a BBH at the end of the simulations.
Figure~\ref{fig:qmtotimbh} shows the total mass of the binaries $M_{\mathrm{tot}}=m_1+m_2$ as a function of the mass ratio $q=m_2/m_1$ (with $m_2<m_1$) of all the IMBH--BH systems. 
The binary with the smallest mass ratio has $q\sim0.04$ and secondary mass $m_2\sim{}4.2$~M$_\odot$. The most massive IMBH--BH binary has masses $m_1\sim438$~\msun{} and $m_2\sim 50$~\msun. 
Values of the mass ratio $q\lesssim0.6$ are the most likely ones, but we find some binaries with $q$ up to  $\sim{}0.99$. Overall, the  IMBH--BH binaries have lower mass ratios than other BBHs (see Figure~5 of \citealt{dicarlo2020b}, where we show that the vast majority of BBHs in young SCs have $q\sim{}0.9-1$). This is expected, because SCs with more than one IMBH are very rare, and IMBHs can thus couple only with lower mass BHs.

We find no mergers of IMBH--BH binaries in our simulations. Adding post-Newtonian terms to our simulations might enhance the chances of IMBH--BH merger during resonant encounters \citep[e.g.,][]{samsing2018,zevin2019,kremer2019}. However, our SCs are relatively small and disrupt quickly, so that these binaries do not have enough time to harden via dynamical interactions. Mergers of BBHs with at least one IMBH component seem to be extremely rare events even in much more massive SCs \citep[e.g.,][]{kremer2020}. 

\begin{figure}
  \center{
    \epsfig{figure=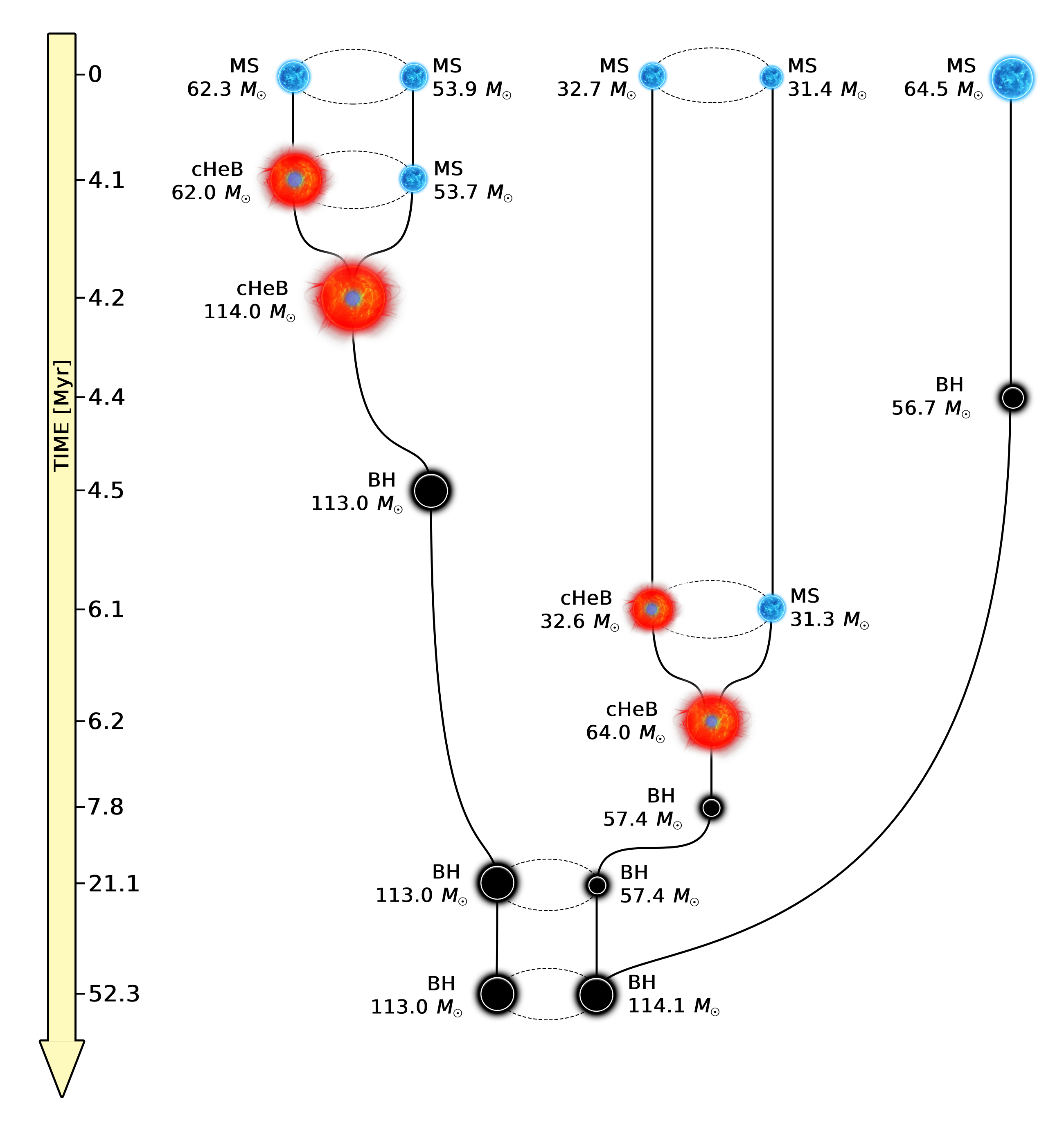,width=8.0cm}
    \caption[Formation history of a binary IMBH found in our simulations]{\label{fig:imbhbin} Formation history of the binary IMBH found in our simulations.  Main sequence stars (with label MS) are represented as blue stars;  core helium burning stars (label cHeB) are visualized as red stars with a blue core; black holes (label BH) are shown as black circles. The mass of each object is shown next to them. The time axis and the size of the objects are not to scale.}}
\end{figure}

We find only one binary composed of two IMBHs (hereafter, binary IMBH), with masses $m_1\sim{113}$~M$_{\odot}$ and $m_2\sim{114}$~M$_{\odot}$. Its formation history is shown in Figure \ref{fig:imbhbin}. First, two massive BHs of $m_1\sim{113}\,\mathrm{M_{\odot}}$ and $m_{\rm a}\sim{57}\,\mathrm{M_{\odot}}$, which formed via runaway collision, dynamically couple and form a BBH. Then, a third BH of mass $m_{\rm b}\sim{57}$~M$_{\odot}$ perturbs the BBH and in the process merges with the secondary BH of the binary, forming the binary IMBH. The formation of this binary involves both the runaway collision and the BBH merger mechanisms.

In our simulations, we do not include relativistic kicks. Were they included, they might have split this binary IMBH at birth. Thus, we calculated a posteriori the relativistic kick induced by the merger of two BHs with $m_{\rm a}=m_{\rm b}=57.4$ M$_\odot$, according to the formulas presented in \cite{maggiorebook}. To calculate the kick, we randomly drew the spin magnitudes of the two BHs from a Maxwellian distribution with $\sigma{}=0.1$  and the spin directions as isotropically distributed over the sphere (see, e.g., \citealt{bouffanais2021} for a discussion of these assumptions). We repeated the calculation of the kick $10^5$ times for different random draws of the spin magnitudes and tilts. We then calculated the probability that the binary system composed of the merger remnant and of the third BH with mass $m_1=113$ M$_\odot $ remains bound after the relativistic kick. We calculate the mass of the merger remnant from \cite{jimenez2017}, yielding $m_{\rm rem}\sim{109}$ M$_\odot$. 
To estimate the orbital properties of the  binary IMBH with masses $m_1$ and $m_{\rm rem}$ we start from an eccentricity $e=0.5$ and a semi-major axis $a=75.5$ AU, which are the values given by {\sc nbody6++gpu} at the time of the merger between $m_{\rm a}$ and $m_{\rm b}$, and we calculate the final semi-major axis and eccentricity (after the relativistic kick) following the equations detailed in the Appendix~A of \cite{hurley2002}. We find that the binary IMBH remains bound in $\sim{}38$\% of our random realizations. If the kicks are drawn from a Maxwellian distribution with $\sigma{}=0.01$ (very small kicks, as predicted by \citealt{fuller2019}), the binary IMBH remains bound in $\sim{}98$\% of the random realizations. The relativistic kick even favours the merger of the final binary IMBH in a few realizations ($\sim{}0.05$\%).

\subsection{Properties of the host SCs}

The presence of an IMBH may influence the evolution and the characteristics of the host SC \citep{mastrobuono2014,bianchini2015,arca2016,askar2017b}. Finding possible signatures of the presence of an IMBH in SCs may help us to understand whether observed SCs host or not an IMBH.

\subsubsection{SC mass}

\begin{figure}
  \center{
    \epsfig{figure=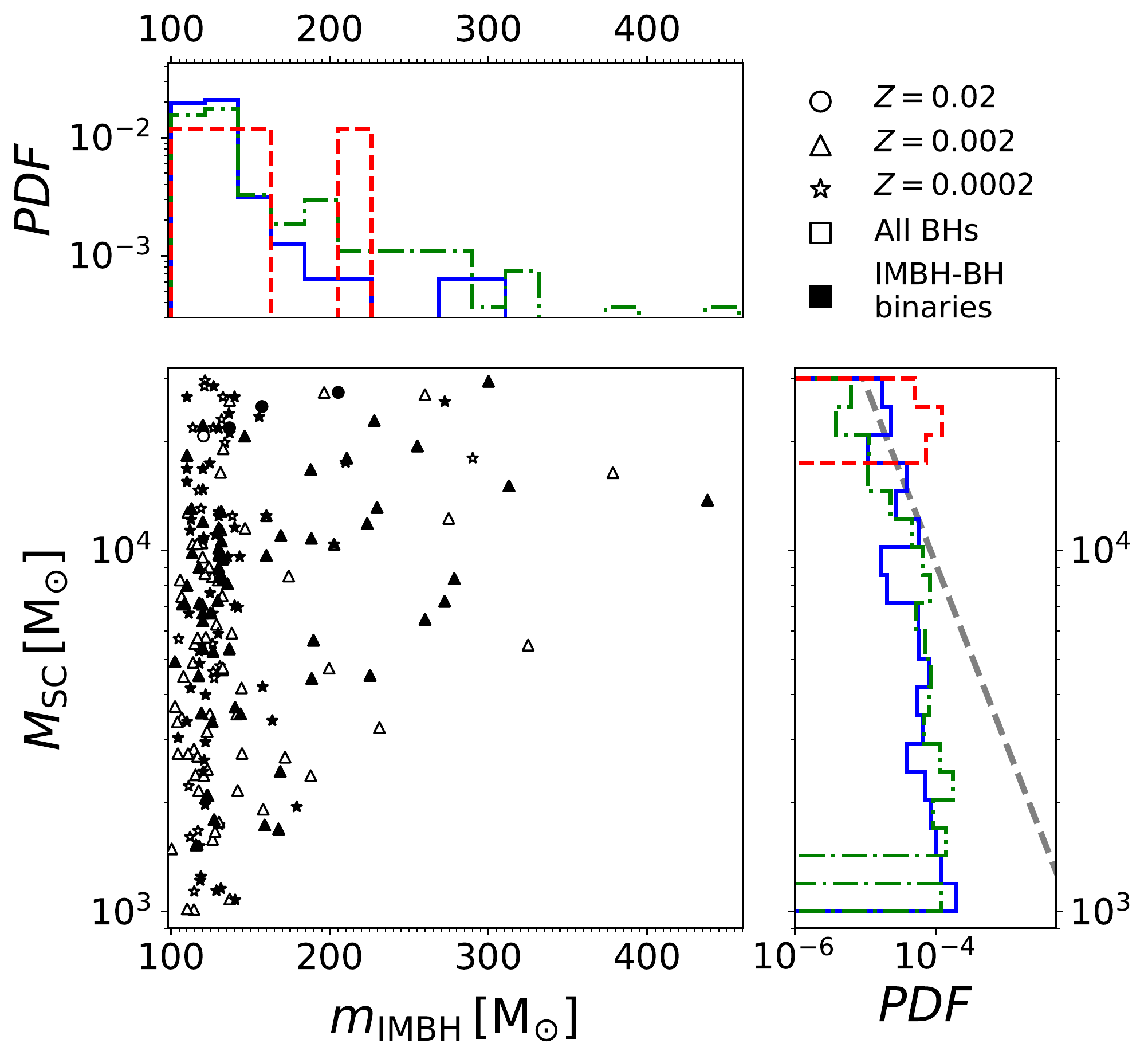,width=8.0cm}
    \caption{{\label{fig:mbhmsc} 
    Mass of the host SC ($M_{\rm SC}$) versus the mass of the IMBH ($m_{\rm IMBH}$). The marginal histograms show the distribution of $M_{\rm SC}$ ($y-$axis) and $m_{\rm IMBH}$ ($x-$axis). The black filled symbols refer to IMBHs in IMBH--BH binaries, while the open symbols are single BHs.
    The solid blue, dot-dashed green and dashed red histograms refer to $Z=0.0002,$ 0.002 and 0.02, respectively. The grey dashed line shows the initial mass function of SCs ($dN/dM_{\rm SC}\propto{}M_{\rm SC}^{-2}$).}}}
\end{figure}

Figure~\ref{fig:mbhmsc} shows the initial mass of the host SC as a function of the mass of the IMBH. The most massive IMBHs form preferentially in massive SCs.
From the marginal histogram which shows the distribution of $M_{\mathrm{SC}}$, it may seem that IMBHs are slightly more numerous in small SCs. However, since we adopted a mass function $dN/dM_{\rm SC}\propto M_{\rm SC}^{-2}$, we simulated many more small SCs than massive ones. As confirmed by Table~\ref{tab:scimbh}, IMBHs form more efficiently in massive SCs where stellar collisions are more effective. For example, IMBHs form in $\sim 8\%$ of the SCs with $10^4\,{\rm M}_{\odot}\leq M_{\rm SC}\leq 3\times{}10^4\,{\rm M}_{\odot}$ and only in $\lesssim 1\%$ of the SCs with $10^3\,{\rm M}_{\odot}\leq M_{\rm SC}< 5\times{}10^3\,{\rm M}_{\odot}$. In our simulations, IMBHs at solar metallicity form only in SCs with mass $\gtrsim{}20500$ M$_\odot$.

\subsubsection{Fractality and initial half mass radius}

\begin{figure}
  \center{
    \epsfig{figure=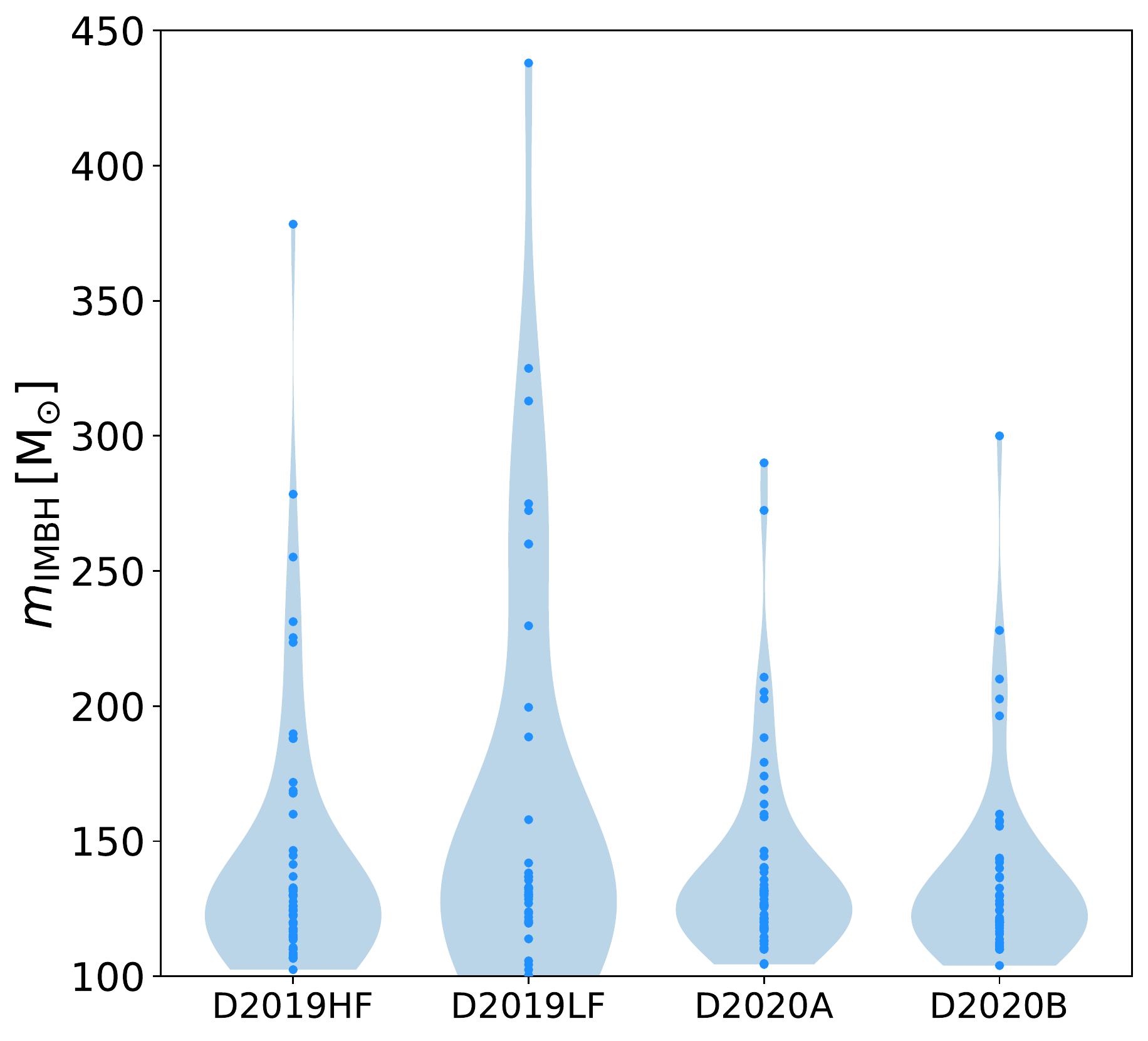,width=8.0cm}
    \caption{\label{fig:m_set_violin} Distributions of IMBH masses for the different simulation sets. Blue filled circles mark the values of the masses, while the horizontal extent of each light blue region (violin plot) is proportional to the number of IMBHs at a given mass value. We refer to Table~\ref{tab:table1} for details on the different sets.}}
\end{figure}

In Figure~\ref{fig:m_set_violin}, we compare the IMBH mass distribution of different simulation sets. Simulations with low fractality (D2019LF) produce more massive IMBHs than high fractality ones (D2019HF). From Table~\ref{tab:scimbh}, however, we see that the D2019HF set is more efficient than D2019LF in producing IMBHs. This means that high fractality helps to produce a larger number of IMBHs, but with lower mass.
This happens because SCs with high fractality are composed of smaller and denser clumps (with a shorter two-body relaxation timescale), where stellar collisions are more likely to occur. On the one hand, this increases their efficiency in forming IMBHs. On the other hand, a small clump does not host many massive stars (in our high-fractality SCs, a single clump may host up to two massive stars at most). Hence, in order to have a runaway collision of massive stars and to form a more massive IMBH, it is necessary that more clumps merge together before the death of their massive stars. In the high-fractality case, the various clumps merge together over a timescale similar or longer than the IMBH formation timescale. While it is easier to form a small IMBH in a single clump, stars from more clumps may be needed to form higher mass IMBHs.

In contrast, low fractality SCs are more similar to monolithic SCs and have a longer two-body relaxation timescale. This means that the runaway collisions will begin later, involving all the massive stars of the SC. On the one hand, the mass of the collision product will not be limited by the mass of the clump, allowing the formation of more massive IMBHs. On the other hand, since the collisions start later, massive stars may die before assembling a sufficiently massive star. This makes low fractality SCs less efficient in forming IMBHs.

We now compare sets D2020A and D2020B in Figure~\ref{fig:m_set_violin} and in Table~\ref{tab:scimbh}. The SCs in D2020A have a smaller initial half-mass radius than in D2020B, but they both have the same degree of fractality ($D=1.6$). This means that the clumps will have approximately the same number of stars, but different densities. As shown in Figure \ref{fig:dens}, the SCs in D2020A have a larger core density in the first $\sim 8\,\rm Myr$, when the runaway collisions take place. Because of this higher initial density, set D2020A is more efficient in forming IMBHs, but the IMBH mass distributions are comparable.

\subsubsection{SC radii}

\begin{figure}
  \center{
    \epsfig{figure=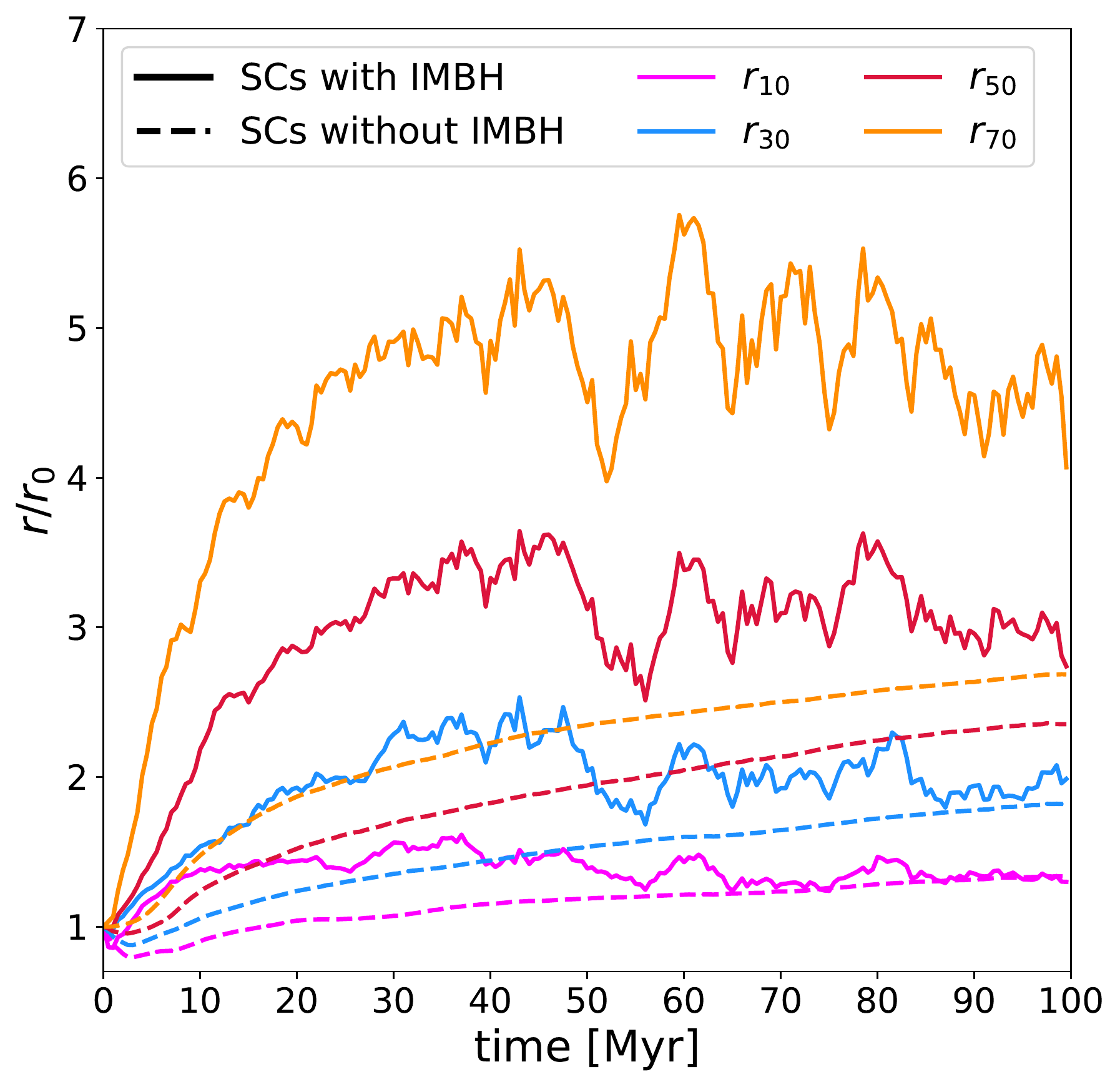,width=8.0cm}
    \caption{{\label{fig:imbhradii} Time evolution of the median of the 10\%, 30\%, 50\% and 70\% Lagrangian radii ($r_{10}$, $r_{30}$, $r_{50}$ and $r_{70}$). Solid lines refer to SCs which contain at least one IMBH; dashed lines refer to SCs which do not contain IMBHs. Each radius is normalized to its initial value $r_0$. A simple moving average over 5 timesteps has been performed to filter out statistical fluctuations.
    }}}
\end{figure}

We check if the presence of an IMBH affects the evolution of the Lagrangian radii of the simulated SCs. Figure~\ref{fig:imbhradii} shows the evolution of the $10\%$, $30\%$, $50\%$ and $70\%$ Lagrangian radii of SCs with and without IMBHs. SCs with IMBHs expand more rapidly in the first few Myrs than SCs without IMBHs. Higher Lagrangian radii expand more, meaning that the expansion is stronger in the outer regions of the SC. 
This effect is a consequence of IMBH dynamics: the IMBH heats the SC, scattering stars to less bound orbits and making the cluster expand \citep{baumgardt2004}.
After the initial expansion, the radii of the SCs with IMBHs flatten out. After $\sim 65$ Myr, the 10\% Lagrangian radii ($r_{10}$) of the SCs with and without IMBH  overlap and start behaving in the same way. At the end of the simulations, the values of the $10\%$ and $30\%$ Lagrangian radii of SCs with and without IMBHs are almost the same. The presence of the IMBH has a stronger impact on Lagrangian radii in the first stages of the evolution of the SCs.

\begin{figure}
  \center{
    \epsfig{figure=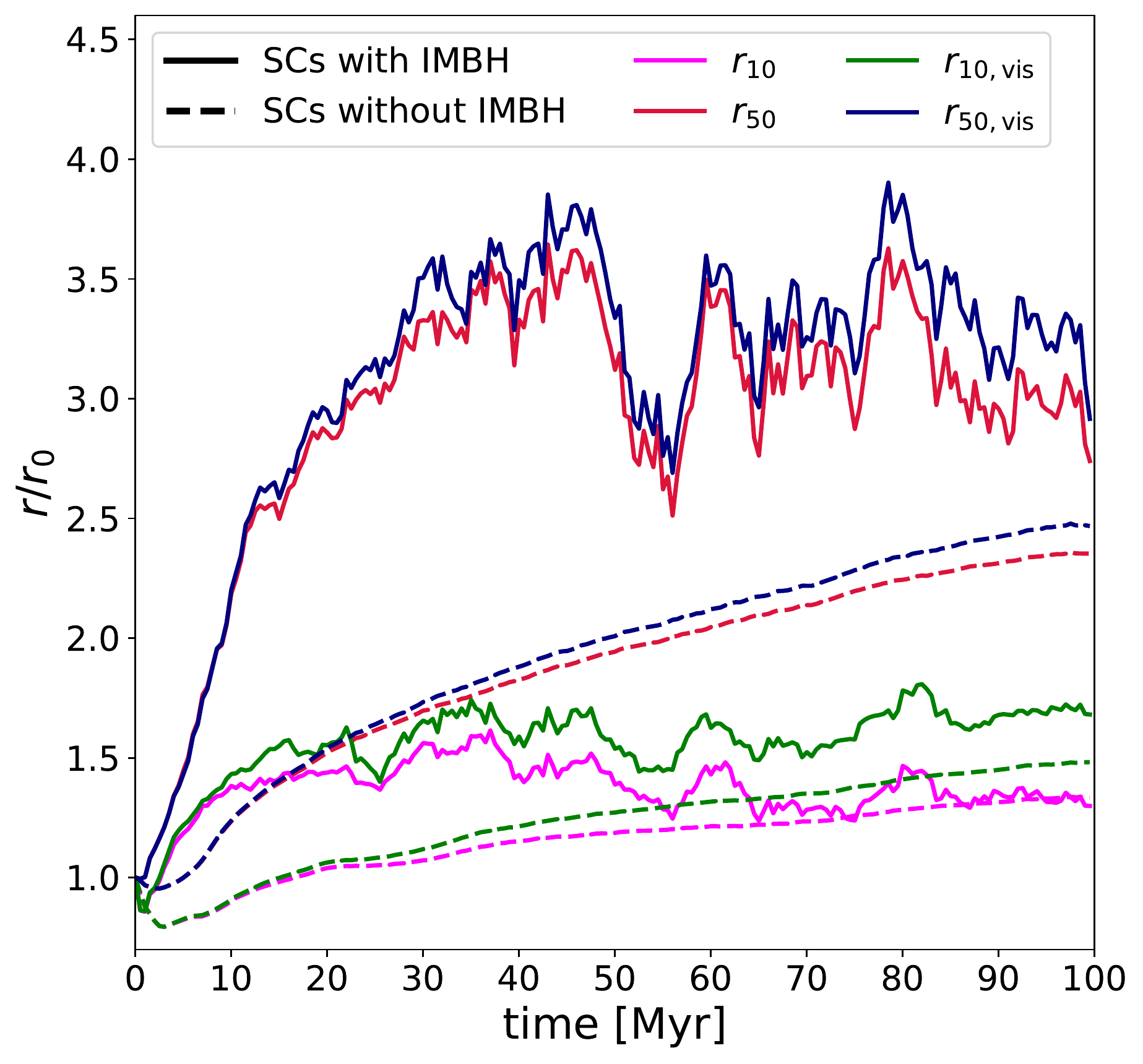,width=8.0cm}
    \caption{{\label{fig:radii_obs} Time evolution of the median of the 10\% and 50\% Lagrangian radii ($r_{10}$ and $r_{50}$) and of the corresponding \emph{visible} Lagrangian radii $r_{\rm 10,\,{}vis}$ and  $r_{\rm 50,\,{}vis}$ (i.e. the 10\% and 50\% Lagrangian radii calculated excluding BHs and neutron stars). Solid lines refer to SCs which contain at least one IMBH; dashed lines refer to SCs without IMBHs. Each radius is normalized to its initial value $r_0$. A simple moving average over 5 timesteps has been performed to filter out statistical fluctuations.
    }}}
\end{figure}

We also check the behavior of Lagrangian radii calculated excluding neutron stars and BHs. In this way, we exclude the dark component and we obtain radii more similar to what observations can tell us, even if we do not account for any other observational bias (for example, we do not remove the lowest mass stars). Hereafter, we call this version of the Lagrangian radii \emph{visible} Lagrangian radii.  Figure~\ref{fig:radii_obs} shows the differences between the  10\% and 50\% Lagrangian radii  and the visible 10\% and 50\% Lagrangian radii. Visible Lagrangian radii are slightly larger than the Lagrangian radii likely due to mass segregation, but the difference does not affect our main result: SCs with IMBHs tend to expand more than SCs without IMBHs.
This result is consistent with the results of previous studies \citep[e.g. ][]{2004ApJ...613.1133B,baumgardt2004,baumgardt2004b,trenti2007,2010ApJ...708.1598T}, which find that the presence of an IMBH in the SC leads to an initial stronger expansion. 

\section{Conclusions}

IMBHs have mass in the range $10^2 - 10^5$ M$_\odot$ and bridge the gap between stellar-sized BHs 
and super-massive BHs. 
Currently, we lack unambiguous evidence of IMBH existence from electromagnetic observations, but we know several strong candidates \citep[e.g.,][]{Farrell2009,godet2014,Reines2015,kiziltan2017,AbbottGW190521,AbbottGW190521astro}. 
Here, we have investigated the formation of IMBHs in young SCs through BBH mergers and the runaway collision mechanism.

In our simulations, 209 IMBHs form via the runaway collision mechanism, while only 9 IMBHs form via BBH mergers. Each SC hosts a maximum of two IMBHs. We find IMBHs  with mass up to $\sim438$~\msun{}, but $\sim78\%$ of all the simulated IMBHs have a mass between $100$~\msun{} and $150$~\msun: less massive IMBHs are more likely to form than the most massive ones.

As expected, IMBH formation is much less efficient at high metallicity, because stellar winds are more effective: only four IMBHs form at solar metallicity. The percentage of the simulated SCs which form at least one IMBH grows as metallicity decreases, going from $0.15\%$ at $Z=0.02$ to $3.5\%$ at $Z=0.0002$.

IMBHs form more efficiently in massive SCs, where dynamics is more important. For example, IMBHs form in $\sim 8\%$ of the SCs with $10^4\,{\rm M}_\odot\leq M_{\rm SC}\leq 3\times{}10^4\,{\rm M}_{\odot}$ and in only $\lesssim 1\%$ of the SCs with $10^3\,{\rm M}_{\odot}\leq M_{\rm SC}< 5\times{}10^3\,{\rm M}_{\odot}$. 

In our simulations, $\sim 54\%$ of all the IMBHs are ejected from their parent SC, preferentially in the first $\sim 25$ Myr from the beginning of the simulations. The IMBHs that remain inside the cluster rapidly sink towards the centre after their formation and stay there until the end of the simulation. IMBHs are more likely to be ejected if their mass is low ($\lesssim{}150$ M$_\odot$) and if the mass of their host SC is large ($>5\times{}10^3$ M$_\odot$).  

Due to their high mass, IMBHs are likely to dynamically interact with other stars and form binary systems. In our simulations, IMBHs spend  $\sim85\%$ of the time, on average, being part of a binary or triple system. IMBHs tend to pair up with other massive BHs in the SC, because dynamical exchanges favour the formation of more massive binaries, which are more energetically stable \citep{hills1980}. 


Overall, our simulations show that IMBHs form quite efficiently in massive young SCs via runaway stellar collisions, especially if the metallicity is relatively low ($Z=0.0002,$ 0.002). While IMBHs are efficient in pairing up dynamically, the occurrence of IMBH--BH mergers is extremely rare ($<1$ every $10^4$ young SCs).

\section*{Acknowledgments}
MM, AB, UNDC, NG, GI and SR acknowledge financial support from the European Research Council (ERC) under European Union's Horizon 2020 research and innovation programme, Grant agreement no. 770017 (DEMOBLACK ERC Consolidator Grant).
MP's contribution to this material is based upon work supported by Tamkeen under the NYU Abu Dhabi Research Institute grant CAP3.
NG acknowledges financial support  by  Leverhulme  Trust  Grant  No.  RPG-2019-350 and Royal Society Grant No. RGS-R2-202004.
\section*{Data Availability}
The data underlying this article will be shared on reasonable request to the corresponding authors.

\bibliography{./undc_imbh}

\appendix\section{Comparing the high-mass tails of the IMBH mass distributions at different metallicities}\label{sec:statapp}

Here, we explore if the mass of the heaviest IMBHs formed in our simulations depends on their progenitors' metallicity. In particular, Figure~\ref{fig:m_z_violin} seems to suggest that the most massive IMBHs form at intermediate metallicity ($Z=0.002\sim{}0.1$ Z$_\odot$) rather than at the lowest considered metallicity ($Z=0.0002\sim{}0.01$ Z$_\odot$). 
To assess this trend with a more quantitative evaluation, we compare the end tail of the IMBH mass distribution at $Z=0.0002$ and $Z=0.002$. 
The mass distribution of all the BHs in the $Z= 0.002$ and the $Z=0.0002$ samples are significantly different, with $p=2.2 \times 10^{-16}$ according to a Kolmogorov-Smirnov (K-S) test. This rises to $p = 0.08$ when only IMBHs (with mass $M > 100$ $M_\odot$), likely because  the K-S test is notoriously insensitive to distribution tails and to small samples \citep[see e.g.][]{10.2307/2240655}.

\begin{figure}
  \center{
    \epsfig{figure=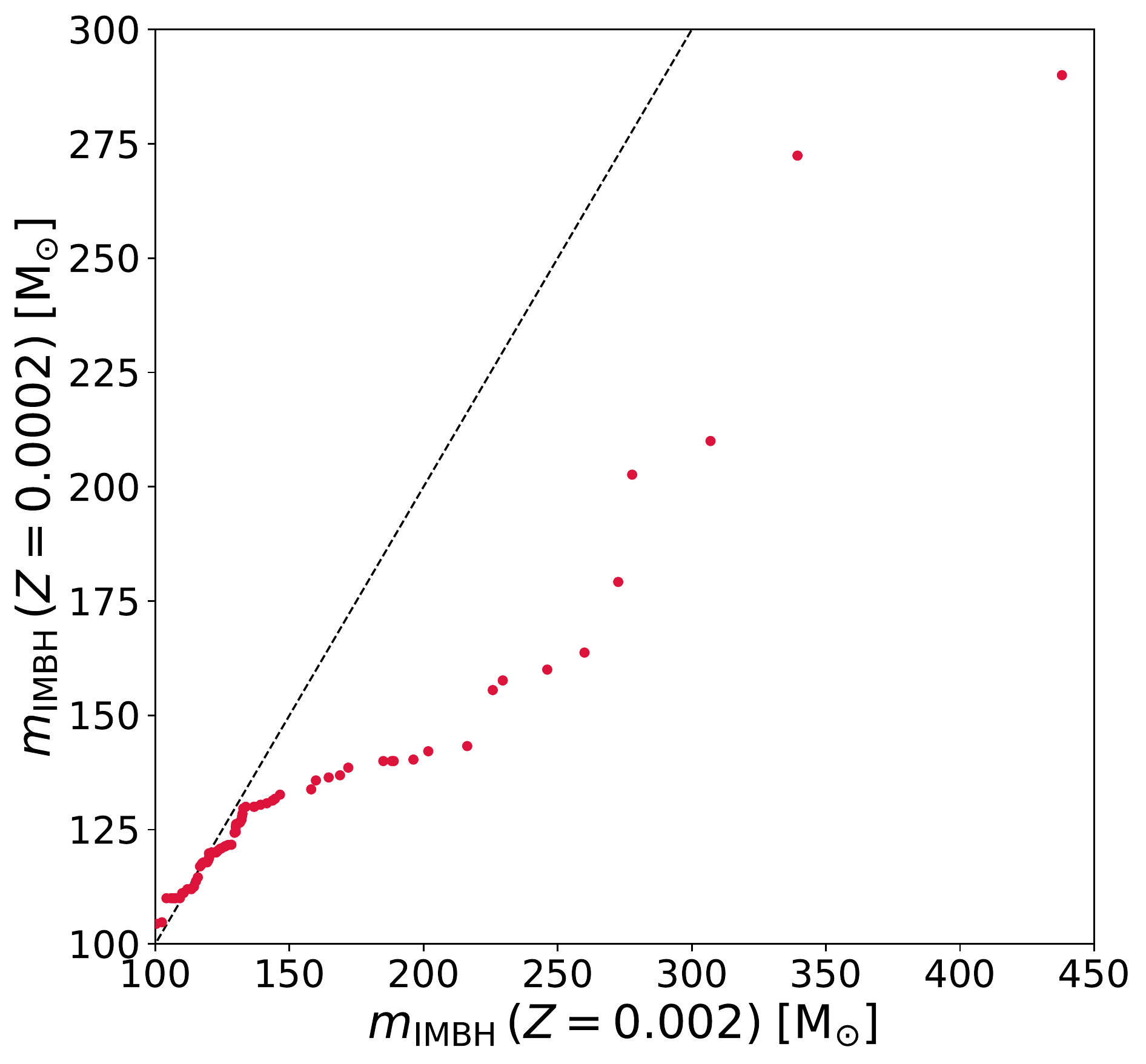,width=8.0cm}
    \caption{\label{fig:ugly} Quantile-quantile plot for the IMBH masses in SCs with $Z = 0.002$ and $Z = 0.0002$. The $x-$ axis shows the values corresponding to selected quantiles of the $Z=0.002$ sample, while the values of the $y-$axis represent the same quantiles for $Z=0.0002$. If the IMBH masses were extracted from the same distribution,  deviations from the diagonal 1:1 black dashed line would be due to random fluctuations only.}}
\end{figure}

Figure~\ref{fig:ugly} shows the quantiles of the IMBH masses for $Z=0.0002$  against those with $Z=0.002$. This quantile-quantile plot allows us to visualize how the two distributions differ: samples drawn from the same distribution would result in points straddling the identity line (black dashed line in Fig.~\ref{fig:ugly}), as each and every quantile would be the same for both samples, modulo random fluctuations. 
This is indeed the case for our two samples in Fig.~\ref{fig:ugly} up until about 130 M$_\odot$, but for higher IMBH masses the $Z=0.002$ simulations appear to produce IMBHs that are systematically heavier than the $Z=0.0002$ simulations. For example, the $90-$th percentile is $228$ M$_\odot$ for the former and $157$ M$_\odot$ for the latter.

It is still far from trivial to conclude whether the high-mass tails and in particular the maximum mass produced is different. 
Comparing maxima is an intrinsically hard problem, because, even for well-behaved distributions, the variance of extremes is much larger than that of, e.g., the mean. A widespread non-parameteric approach for testing differences in maxima relies on an exact Fisher test \citep[][]{fisher1992statistical} applied to the contingency table obtained by counting how many BHs lie above selected mass cut-offs in each of the two metallicity samples. This is similar to applying a $\chi ^2$ test, but is exact for the case in which cell counts are very low, as we expect them to be for large enough mass cut-offs.
We find that the null hypothesis (that both metallicities produce IMBHs above and below each mass cutoff in the same proportion) cannot be rejected for cutoffs at the $90$th, $95$th, and $99$th percentiles of the joined samples, so we conclude that at the moment our data do not constrain the impact of metallicity on maximum IMBH mass, at least if we forego making assumptions on the underlying parent distribution.

A general concern with this comparison could be that the number of simulations run at the two different metallicities is different, so the samples can be compared fairly only through a bootstrap re-sampling that takes this difference into account. We thus repeated our analysis by randomly re-sampling from the $Z=0.002$ sample either a number of IMBHs equal to that in the less numerous $Z=0.0002$ sample or a number of IMBHs corresponding to an equivalent number of simulations. Even with this test the results from the above still hold, i.e. no significant difference in the high-mass tails of the distributions could be ascertained non-parametrically.

%

\end{document}